\documentclass{article}
\pdfoutput=1
\usepackage[margin=1.25in]{geometry}
\usepackage{braket}
\usepackage{amsmath}
\usepackage{graphicx}
\usepackage{xcolor}
\usepackage{authblk}
\usepackage{lmodern}
\usepackage[sorting=none]{biblatex}
\usepackage{amssymb}
\usepackage{tikz}
\usetikzlibrary{positioning}
\usetikzlibrary{shapes.geometric}

\title{Mutual Information Scaling for Tensor Network Machine Learning}

\author[1,3]{Ian Convy\thanks{ian\_convy@berkeley.edu}}
\author[1,3]{William Huggins\thanks{Current Affiliation: Google Quantum AI, Mountain View, CA, USA}}
\author[2,3]{Haoran Liao}
\author[1,3]{K. Birgitta Whaley}

\affil[1]{Department of Chemistry, University of California, Berkeley, CA 94720, USA}
\affil[2]{Department of Physics, University of California, Berkeley, CA 94720, USA}
\affil[3]{Berkeley Quantum Information and Computation Center, University of California, Berkeley, CA 94720, USA}

\date{}

\begin{document}

\maketitle

\begin{abstract}
Tensor networks have emerged as promising tools for machine learning, inspired by their widespread use as variational ansatze  in quantum many-body physics. It is well known that the success of a given tensor network ansatz depends in part on how well it can reproduce the underlying entanglement structure of the target state, with different network designs favoring different scaling patterns. We demonstrate here how a related correlation analysis can be applied to tensor network machine learning, and explore whether classical data possess correlation scaling patterns similar to those found in quantum states which might indicate the best network to use for a given dataset. We utilize mutual information as measure of correlations in classical data, and show that it can serve as a lower-bound on the entanglement needed for a probabilistic tensor network classifier. We then develop a logistic regression algorithm to estimate the mutual information between bipartitions of data features, and verify its accuracy on a set of Gaussian distributions designed to mimic different correlation patterns. Using this algorithm, we characterize the scaling patterns in the MNIST and Tiny Images datasets, and find clear evidence of boundary-law scaling in the latter. This quantum-inspired classical analysis offers insight into the design of tensor networks which are best suited for specific learning tasks.
\end{abstract}

\section{Introduction}

Tensor decompositions~\cite{Kolda_Bader_2009}\cite{Hackbusch_2012}, often represented graphically as \textit{tensor networks}~\cite{Bridgeman_Chubb_2017}, have proven to be useful for analyzing and manipulating vectors in very high-dimensional spaces. One area of particular interest has been the application of tensor network methods to quantum many-body physics~\cite{Eisert_2013}\cite{Verstraete_Cirac_Murg_2008}, where the network serves as a parameterized ansatz that can be variationally optimized to find the ground state of a target Hamiltonian~\cite{White_1992} or simulate quantum dynamics~\cite{Vidal_2003}. Inspired by these successes in quantum physics, there has been an increased focus in applying tensor networks to machine learning~\cite{Cohen_Sharir_Shashua_2016}\cite{Stoudenmire_Schwab_2016}\cite{Huggins_Patil_Mitchell_Whaley_Stoudenmire_2019}, where the learning problem is formulated as linear regression on a massively expanded feature space. Although this subset of machine learning research  is relatively new, tensor network approaches for classification have already yielded promising results on common benchmark datasets.~\cite{Stoudenmire_Schwab_2016}\cite{Stoudenmire_2018}\cite{Cheng_Wang_Zhang_2020} 

A central question that arises wherever tensor networks are used, be it in quantum many-body physics or machine learning, is which network structure to choose for a given task. Since matrix product states (MPS) serve as the underlying ansatz for the highly successful DMRG algorithm used to calculate ground-state energies~\cite{Schollwock_2011}, researchers in quantum many-body physics have worked to understand the strengths and limitations of these networks. Ultimately, the success of DMRG in 1-D systems is made possible by the short-range interactions present in many Hamiltonians, which result in ground states that possess exponentially decaying correlations and localized entanglement that obeys an ``area law" or more properly a \textit{boundary law}~\cite{Eisert_Cramer_Plenio_2010}. These discoveries have helped motivate the development of other network structures such as projected entangled pair states (PEPS)~\cite{Orus_2014} and the multiscale entanglement renormalization ansatz (MERA)~\cite{Vidal_2008} to deal with multidimensional lattices and quantum critical points respectively.

The purpose of our work is to take the entanglement scaling analysis that has been so illuminating in quantum many-body physics, and adapt it for use on the classical data commonly found in machine learning. Through this analysis, we seek to understand which tensor networks would be most appropriate for specific learning tasks. The body of the paper is organized into five parts: Section \ref{sec tensor-networks} begins with an overview of tensor networks and their application to machine learning. In Section \ref{sec correlation-scaling} we review how entanglement scaling relates to tensor network methods in quantum many-body physics, and then extend this analysis to classical data by using the mutual information (MI), which provides a generalized measure of correlation. We show that when using tensor networks for probabilistic classification of orthogonal inputs, the MI of the data provides a lower-bound on the entanglement and thus the connectivity of the tensors. Section \ref{sec mi-estimation} introduces a numerical method for estimating the MI of a dataset given access to only a finite number of samples. In Section \ref{sec GMRF}, we test the accuracy of this method on a set of Gaussian distributions engineered to have different MI scaling patterns with respect to spatial partitioning of the variables. In Section \ref{sec real-data} we estimate the MI scaling of MNIST~\cite{MNIST} and the Tiny Images~\cite{Torralba_Fergus_Freeman_2008}, two well-known image datasets commonly used in machine learning, and find evidence that the MI between a centered, square patch of pixels and the surrounding pixels scales with the boundary of the inner patch (a boundary law), rather than with the number of pixels (a volume law). This boundary-law scaling suggests that networks with an underlying 2-D grid structure such as PEPS would be especially well-suited for machine learning on images.

\section{Tensor Networks}\label{sec tensor-networks}

\subsection{Fundamentals}

For the purposes of this work, a \textit{tensor} is an array of numbers with a finite number of indices $n$, each denoted by a distinct subscript. The value of $n$ is called the \textit{order} of the tensor, meaning that vector $v_i$ is a first-order tensor, matrix $M_{ij}$ is a second-order tensor, $A_{ijk}$ is a third-order tensor, and so on. The goal of a tensor \textit{network} is to represent a higher-order tensor as the contraction of a set of lower-order tensors. Since the number of elements in a tensor scales exponentially with the order, a tensor network representation using lower-order tensors can contain exponentially fewer elements than the original tensor, and thus significantly reduce the amount of computational resources required for numerical analysis. For example, a non-cyclic or \textit{open} MPS network (also called a \textit{tensor train decomposition}~\cite{Oseledets_2011}) represents the $n$th-order tensor $C_{i_1i_2...i_n}$ as the contraction of a sequence of matrices and third-order tensors
\begin{equation}\label{eq1}
C_{i_1i_2...i_n} = \sum_{\alpha_1...\alpha_{n-1}}M^{(1)}_{i_1\alpha_1}A^{(2)}_{\alpha_1i_2\alpha_2}A^{(3)}_{\alpha_2i_3\alpha_3}...M^{(n)}_{\alpha_{n-1}i_n},    
\end{equation}
where $i_1,...,i_n$ are the indices corresponding to the higher-order tensor $C$ and $\alpha_1...\alpha_{n-1}$ are the internal indices of the network that get contracted together. If the indices $i_j$ of $C$ each have dimension $d$ and the internal indices $\alpha_j$ each have dimension $m$, then the number of elements in the MPS is $\mathcal{O}(dm^2n)$ while $C$ has $d^n$ elements. If the internal dimension is chosen such that $m \ll d^{\frac{n}{2}}$, the memory resources needed to represent $C$ are greatly reduced. This efficiency typically comes at the cost of accuracy, however, since most higher-order tensors cannot be exactly represented by a reasonably-sized MPS or other tensor network, and thus some approximation error is introduced. 

When working with tensor networks, it is common to represent expressions such as Eq.~\eqref{eq1} using a graphical notation, where the tensors are represented as geometric shapes and the indices are represented as lines or \textit{legs} protruding outward~\cite{Biamonte_Bergholm_2017}\cite{Orus_2014}\cite{Biamonte_2020}. The contraction of a pair of indices between two tensors is expressed by connecting the legs of the two tensors together. For example, an open MPS can be expressed graphically as a 1-D chain, as shown in Figure~\ref{fig:mps_diagram}, with the legs of neighboring tensors connected together. A major advantage of the graphical notation is that patterns of connectivity are very clear, even in contractions that involve a large number of tensors. In this paper we augment our tensor network equations with these diagrams to make them easier to visualize.   

\begin{figure}
    \centering
    \begin{equation*}
        \sum_{\alpha_1...\alpha_4}M^{(1)}_{i_1\alpha_1}A^{(2)}_{\alpha_1i_2\alpha_2}A^{(3)}_{\alpha_2i_3\alpha_3}A^{(4)}_{\alpha_3i_4\alpha_4}M^{(5)}_{\alpha_4i_5}
    \end{equation*}
    \begin{tikzpicture} [
        tensor/.style={circle, draw, fill=gray!40, minimum size=15}
        ]
        \node[tensor] (1) {};
        \draw[black, thick] (1.south){}+(0, -0.5) -- (1.south);
        \foreach \x [count=\i] in {2,...,5} {
            \node[tensor] (\x) [right=0.5 of \i] {};
            \draw[black, thick] (\i.east) -- (\x.west);
            \draw[black, thick] (\x.south){}+(0, -0.5) -- (\x.south);
        }
    \end{tikzpicture}
    \caption{Tensor network diagram for an open MPS containing five tensors, with the corresponding equation given above. The first and last tensors are matrices labeled as $M$, with the rest being third-order tensors labeled generically as $A$. The contractions between neighboring tensors are made explicit in the graphical notation, without needing to keep track of specific indices. The order of the MPS is also clear from the number of uncontracted legs.}
    \label{fig:mps_diagram}
\end{figure}

\subsection{Tensor Networks for Machine Learning}\label{sec tensor-network-ml}

The most common forms of discriminative tensor network machine learning can be understood in terms of linear regression, where the model output is generated by a weighted sum of the inputs~\cite{Seber_Lee_2012}. This regression is performed by computing the inner product between a feature vector $\vec{x}$ representing the data and a weight vector $\vec{w}$ representing the model. Taken together with an additive bias $b$, these produce a scalar output $y$ that lies along a hyperplane
\begin{equation}
    y = \sum_i w_ix_i + b.
\end{equation}
To make these models more expressive, it is common to first  transform $\vec{x}$ using a set of feature maps, and then perform the regression. The advantage of such a transformation is that while the output of the model is still linear in the transformed space, it can be a highly non-linear function of the input when mapped back to the original space. A significant drawback is that the outputs of the feature maps may be very high-dimensional and thus too large to store or manipulate. This problem is most often solved using the \textit{kernel trick}~\cite{Hastie_Tibshirani_Friedman_2009}, where the inner product between feature mappings is used for regression rather than the full feature map vectors. Unfortunately, since the computational cost of many kernel trick methods scales quadratically with the number of samples, this can be impractical for large datasets.

Tensor networks offer a different solution. First, the feature  map is constrained to have a tensor product structure
\begin{equation}\label{eq prod_state_feat}
X_{i_1i_2...i_k} = \bigotimes^k_{j=1}f_{i_j}(\vec{x}) \enspace = \enspace \tikz[baseline=0.1ex, data/.style={draw, fill=gray!40, minimum size=7}] {
        \node[data] (1) {};
        \draw[black, thick] (1.north){}+(0, 0.25) -- (1.north);
        \foreach \x [count=\i] in {2,3} {
            \node[data] (\x) [right=0.25 of \i] {};
            \draw[black, thick] (\x.north){}+(0, 0.25) -- (\x.north);
        }
}\enspace ... \enspace \tikz[baseline=0.1ex, data/.style={draw, fill=gray!40, minimum size=7}] {
        \node[data] (4) {};
        \draw[black, thick] (4.north){}+(0, 0.25) -- (4.north);
} \enspace,    
\end{equation}
where $X_{i_1i_2...i_k}$ is the transformed representation of $\vec{x}$, and $\{f_{i_j}\}$ are vector-valued functions transforming the original features. Next, regression is performed directly on the transformed features by fully contracting $X_{i_1i_2...i_k}$ with the weight tensor $W_{i_1i_2...i_k}$ which is represented by a tensor network. For example, choosing $W_{i_1i_2...i_k}$ to be an MPS gives
\begin{equation}
\begin{split}
    y &= \sum_{i_1...i_k}W_{i_1...i_k}X_{i_1...i_k}
    \\
    &= \sum_{i_j,\alpha_j}M^{(1)}_{i_1\alpha_1}A^{(2)}_{\alpha_1i_2\alpha_2}...M^{(k)}_{\alpha_{k-1}i_k}\bigotimes^k_{j=1}f_{i_j}(\vec{x})
    \\
    &= \sum_{i_j,\alpha_j} f_{i_1}(\vec{x}) M^{(1)}_{i_1\alpha_1}f_{i_2}(\vec{x})A^{(2)}_{\alpha_1i_2\alpha_{2}}...f_{i_k}(\vec{x})M^{(k)}_{\alpha_{k-1}i_k} \\
    &= \tikz[baseline=1.3ex, data/.style={draw, fill=gray!40, minimum size=7}, tensor/.style={draw, circle, fill=gray!40, minimum size=7}] {
        \node[data] (1_data) {};
        \node[tensor] (1_mps) [above=0.25 of 1_data] {};
        \draw[black, thick] (1_data.north) -- (1_mps.south);
        \node[data] (2_data) [right=0.25 of 1_data] {};
        \node[tensor] (2_mps) [above=0.25 of 2_data] {};
        \draw[black, thick] (1_mps.east) -- (2_mps.west);
        \draw[black, thick] (2_data.north) -- (2_mps.south);
        \node[data] (3_data) [right=0.25 of 2_data] {};
        \node[tensor] (3_mps) [above=0.25 of 3_data] {};
        \draw[black, thick] (2_mps.east) -- (3_mps.west);
        \draw[black, thick] (3_data.north) -- (3_mps.south);
        \draw[black, thick] (3_mps.east){}+(0.25, 0) -- (3_mps.east);
    } \enspace ... \enspace \tikz[baseline=1.3ex, data/.style={draw, fill=gray!40, minimum size=7}, tensor/.style={draw, circle, fill=gray!40, minimum size=7}] {
        \node[data] (4_data) {};
        \node[tensor] (4_mps) [above=0.25 of 4_data] {};
        \draw[black, thick] (4_data.north) -- (4_mps.south);
        \draw[black, thick] (4_mps.west){}+(-0.25, 0) -- (4_mps.west);
    }
\end{split}
\end{equation}
where the last sum can be performed efficiently by contracting the tensors from left to right. For large $k$ a tensor network representation is essential, since the raw weight tensor $W_{i_1...i_k}$ has far too many elements to operate on directly. However, it is not obvious which type of network to use. Although MPS networks are the most commonly used in the literature, tree tensor networks (TNN)~\cite{Shi_Duan_Vidal_2006} and MERA have also been employed~\cite{Liu_Ran_Wittek_Peng_Garcia_Su_Lewenstein_2019}\cite{Reyes_Stoudenmire_2021}\cite{Cong_Choi_Lukin_2019}. The wide variety of possible tensor networks raises an obvious question: which structure is best suited for a given machine learning task? In the next section we first describe how the many-body physics community has used the spatial scaling patterns of quantum correlations to answer a similar question when modeling quantum states, and then adapt this analysis for machine learning.  

\section{Correlation Scaling}\label{sec correlation-scaling}

\subsection{Entanglement Scaling in Quantum Systems}\label{sec correlation-scaling-qm}

Entanglement is a defining property of quantum mechanics~\cite{Schrodinger_1935}, and is the source of all correlations between components of a pure-state composite system~\cite{Plenio_Virmani_2005}. Although there are multiple methods of quantifying entanglement, the \textit{entropy of entanglement} is a widely used measure for entanglement between bipartitions of a composite system. For a pure state defined by the joint density matrix $\rho_{AB}$ with reduced density matrices $\rho_A$ and $\rho_B$ corresponding to the bipartitions $A$ and $B$, the entanglement entropy is defined as the von Neumann entropy of $\rho_A$ (or equivalently $\rho_B$)
\begin{equation}\label{eq5}
    E(A, B) = -\text{Tr}(\rho_A\log\rho_A).
\end{equation}
A connection between the entanglement entropy of a quantum state and its structure can be made using the \textit{Schmidt decomposition}~\cite{Ekert_Knight_1995}, which is defined for state $\ket{\psi}$ on the combined Hilbert space $\mathcal{H}_A \otimes \mathcal{H}_B$ as
\begin{equation}\label{eq6}
    \ket{\psi} = \sum^r_{\alpha=1} \lambda_\alpha \ket{s^A_\alpha}\ket{s^B_\alpha},
\end{equation}
where $r$ is the Schmidt rank, the $\lambda_\alpha$ are the Schmidt coefficients, and $\ket{s^A_\alpha}, \ket{s^B_\alpha}$ are the orthonormal Schmidt basis states in $\mathcal{H}_A$ and $\mathcal{H}_B$ respectively. Substituting Eq.~\eqref{eq6} into Eq.~\eqref{eq5} gives an expression for the entanglement in terms of the Schmidt coefficients
\begin{equation}\label{eq7}
    E(A, B) = -\sum^r_{\alpha=1}|\lambda_\alpha|^2\log(|\lambda_\alpha|^2).
\end{equation}
Formally, the Schmidt decomposition may be regarded as a singular value decomposition (SVD) of the matrix $C$ of coefficients that form $\ket{\psi}$:
\begin{equation}
\begin{split}
    \ket{\psi} &= \sum_{ij}C_{ij}\ket{i_A}\ket{j_B}
    \\
    &= \sum_{ij\alpha_1\alpha_2} V_{i\alpha_1} \Lambda_{\alpha_1, \alpha_2} U^{\dagger}_{\alpha_2, j}\ket{i_A}\ket{j_B}
    \\
    &= \sum_{ij\alpha} \lambda_\alpha V_{i\alpha}\ket{i_A}U_{j,\alpha}\ket{j_B}
    \\
    &= \sum_\alpha \lambda_\alpha
    \ket{s^A_\alpha}\ket{s^B_\alpha}, 
\end{split}
\end{equation}
where the rows of $C$ correspond to the computational basis states $\ket{i_A}$ in $\mathcal{H}_A$ and the columns correspond to the computational basis states $\ket{j_B}$ in $\mathcal{H}_B$. The diagonal matrix $\Lambda$ can be truncated so that it contains only the non-zero singular values of $C$, which are then equal to the Schmidt coefficients $\lambda_\alpha$. Whenever there is more than one non-zero $\lambda_\alpha$, the state possesses some degree of entanglement. Since the Schmidt decomposition is an SVD, the set of $\lambda_\alpha$ is guaranteed to be unique, and the Schmidt rank will be minimized with respect to all possible basis sets. Using the SVD matrices explicitly, we can write the Schmidt decomposition as a small tensor network
\begin{equation}\label{eq svd}
    \ket{\psi} = \sum_{i, j, \alpha_1, \alpha_2} V_{i\alpha_1} \Lambda_{\alpha_1, \alpha_2} U^{\dagger}_{\alpha_2, j}\ket{i_A}\ket{j_B} \enspace \rightarrow \enspace \tikz[baseline=-1ex, unitary/.style={circle, draw, fill=gray!40, minimum size=7}, diag/.style={draw, diamond, fill=gray!40, minimum size=7}] {
        \node[unitary] (v) {};
        \node[diag] (d) [right=0.25 of v] {};
        \node[unitary] (u) [right=0.25 of d] {};
        \draw[black, thick] (v.south){}+(0,-0.25) -- (v.south);
        \draw[black, thick] (v.east) -- (d.west);
        \draw[black, thick] (d.east) -- (u.west);
        \draw[black, thick] (u.south){}+(0,-0.25) -- (u.south);
    } \enspace ,
\end{equation}
where $V$, $U$ are unitary matrices that map the basis states $\ket{i_A}$, $\ket{j_B}$ to the Schmidt bases of $\mathcal{H}_A$ and $\mathcal{H}_B$ respectively. It is important to note that this mathematical description of entanglement, which is based on the singular values, can be used to characterize a tensor regardless of whether it represents a truly quantum object.

The fact that Eq.~\eqref{eq7} arises from a Schmidt decomposition is key to understanding the entanglement scaling properties of tensor networks. As a simple example, the (open) MPS representation of an $N$-component quantum system used in algorithms such as DMRG is given by a contraction of second- and third-order tensors, each corresponding to a physical degree of freedom
\begin{equation}\label{eq8}
    \ket{\psi} = \sum_{i_j, \alpha_j} M^{(1)}_{i_1\alpha_1}A^{(2)}_{\alpha_1i_2\alpha_2}\cdots M^{(N)}_{\alpha_{N-1}i_N}\ket{i_1}\cdots\ket{i_N} \rightarrow \enspace \tikz[baseline=2.3ex, tensor/.style={draw, circle, fill=gray!40, minimum size=7}] {
        \node[tensor] (1_mps) [above=0.25 of 1_data] {};
        \draw[black, thick] (1_mps.south){}+(0,-0.25) -- (1_mps.south);
        \node[tensor] (2_mps) [above=0.25 of 2_data] {};
        \draw[black, thick] (1_mps.east) -- (2_mps.west);
        \draw[black, thick] (2_mps.south){}+(0,-0.25) -- (2_mps.south);
        \node[tensor] (3_mps) [above=0.25 of 3_data] {};
        \draw[black, thick] (2_mps.east) -- (3_mps.west);
        \draw[black, thick] (3_mps.south){}+(0,-0.25) -- (3_mps.south);
        \draw[black, thick] (3_mps.east){}+(0.25, 0) -- (3_mps.east);
    } \enspace ... \enspace \tikz[baseline=2.3ex, tensor/.style={draw, circle, fill=gray!40, minimum size=7}] {
        \node[tensor] (4_mps) [above=0.25 of 4_data] {};
        \draw[black, thick] (4_mps.south){}+(0,-0.25) -- (4_mps.south);
        \draw[black, thick] (4_mps.west){}+(-0.25, 0) -- (4_mps.west);
    }.
\end{equation}
If the physical indices are grouped together into two contiguous partitions $A$ and $B$, with the internal indices contracted within each partition, then Eq.~\eqref{eq8} can be rewritten as
\begin{equation}\label{eq9}
    \ket{\psi} = \sum_{i, j, \alpha} M^{(A)}_{i\alpha}M^{(B)}_{\alpha j}\ket{i_A}\ket{j_B} \enspace \rightarrow \enspace 
    \tikz[baseline=-1.3ex, tensor/.style={circle, draw, fill=gray!40, minimum size=7}] {
        \node[tensor] (1) {};
        \node[tensor] (2) [right=0.25 of 1] {};
        \draw[black, thick] (1.south){}+(0, -0.25) -- (1.south);
        \draw[black, thick] (1.east) -- (2.west);
        \draw[black, thick] (2.south){}+(0,-0.25) -- (2.south);
    } \enspace ,
\end{equation}
where $i$ and $j$ are the combined physical indices of partition $A$ and partition $B$ respectively. If the dimension of index $\alpha$ is $m$, then Eq.~\eqref{eq9} is a canonical decomposition~\cite{Hitchcock_1927} with $m$ terms, having a form similar to that of the SVD in Eq.~\eqref{eq svd}. Since the SVD, and therefore the Schmidt decomposition, represents the canonical decomposition with the fewest possible terms, the Schmidt rank of an MPS is always upper bounded by $m$. Through Eq.~\eqref{eq7}, this implies that the entanglement entropy represented by an MPS of bond dimension $m$ is bounded by
\begin{equation}\label{eq10}
    E_{MPS} \leq \log(m), 
\end{equation}
where the inequality is saturated if $m$ is equal to the Schmidt rank and if the singular values are all $m^{-1}$.

This analysis can be extended beyond MPS~\cite{Evenbly_Vidal_2011}, with the index $\alpha$ representing the combination of all indices connecting the tensors in the two partitions. Assuming a maximum bond dimension given by $m$ and a number of connecting indices $n$, the dimension of $\alpha$ is $m^n$ and therefore Eq.~\eqref{eq10} can be extended to a general tensor network as 
\begin{equation}\label{eq network entangle}
    E_{TN} \leq n\log(m).
\end{equation}
Assuming a fixed bond dimension $m$, differences in entanglement scaling between tensor networks arise from differences in the value of $n$, which depends on the geometry of the network. For tensor networks which conform to the physical geometry of the composite system, such as MPS for 1-D systems and PEPS for 2-D systems, the number of indices connecting two partitions is determined by the size of the interface between the partitions. Given a simple partitioning of the system into a contiguous, hypercubic patch of length $L$ and the surrounding outer patch, the interface scales with the boundary of the inner patch. If the physical lattice dimension is $d$, the entanglement follows a boundary-law scaling expression
\begin{equation}\label{eq boundary law}
    E_{TN} \leq 2dL^{d-1}\log(m) = \mathcal{O}(L^{d-1}).
\end{equation}
This scaling behavior stands in sharp contrast to that of a random quantum state, whose entanglement will scale with the total size of the inner patch~\cite{Page_1993} rather than its boundary in what is sometimes referred to as a ``volume law". The success of methods like DMRG is only possible because the ground states of common Hamiltonians do not resemble states that have been randomly sampled from the Hilbert space, but instead tend to possess localized, boundary law entanglement that can be readily captured with the MPS ansatz. The existence of such scaling patterns has been proven for the ground states of 1-D gapped quantum systems~\cite{Hastings_2007}, and for harmonic lattice systems of arbitrary dimension~\cite{Cramer_Eisert_Plenio_Dreibig_2006}. They have also been conjectured to exist in the ground states of most local, gapped quantum systems regardless of dimension~\cite{Eisert_Cramer_Plenio_2010}. Different tensor networks need to be employed when the ground state is suspected to violate the strict boundary law, with networks such as MERA being used to handle the $\log(L)$ corrections found in many critical-phase Hamiltonians~\cite{Vidal_Latorre_Rico_Kitaev_2003}. In any case, the ultimate goal of these tensor network ansatzes is to match the known or predicted entanglement scaling of the quantum state with the entanglement scaling of the network.  

\subsection{Correlations in Classical Data}\label{sec correlation-scaling-mi}

The preceding analysis used entanglement to quantify correlations in a system that was explicitly quantum mechanical. To carry out a similar analysis on classical data, we desire a more general quantity. A reasonable candidate is the \textit{mutual information} (MI)~\cite{Ebrahimi_Soofi_Soyer_2010}, defined as
\begin{equation}\label{eq13}
    I(A : B) = S(A) + S(B) - S(AB),
\end{equation}
where $S$ is the entropy of the probability distributions associated with marginal variables $A$, $B$ and the joint variable $AB$. Qualitatively, the MI describes the amount of information we gain about one variable when we learn the state of the other, offering the most general measure of correlation. The MI can be calculated for either quantum or classical data, depending on whether the von Neumann or Shannon entropies are used. For a pure quantum state $S(AB) = 0$, and therefore the MI is equal to twice the entanglement.

An alternative but equivalent representation of the MI, which we make use of in Sec.~\ref{sec mi-estimation}, comes from the \textit{Kullback-Liebler divergence} (KL-divergence), which is defined for two discrete probability distributions $P$ and $Q$ on variable space $\mathcal{X}$ as 
\begin{equation}
    D_{KL}(P||Q) = \sum_{x \in \mathcal{X}}P(x)\log\frac{P(x)}{Q(x)},
\end{equation}
with an analogous definition for continuous variables that replaces the sum with an integral over probability densities. For a joint probability distribution $P$ over variables $A$ and $B$ in spaces $\mathcal{A}$ and $\mathcal{B}$, the MI is equal to the KL-divergence between the joint distribution $P(A, B)$ and the uncorrelated product-of-marginals distribution $P(A)P(B)$, i.e.
\begin{equation}\label{eq mi-kl}
    I(A : B) = \sum_{a \in \mathcal{A}}\sum_{b \in \mathcal{B}}P(a, b)\log\frac{P(a, b)}{P(a)P(b)}.
\end{equation}
While not formally a metric, the KL-divergence can be viewed as measuring the distance between two distributions, so Eq.~\eqref{eq mi-kl} represents the MI as the distance between $P(A, B)$ and the uncorrelated distribution $P(A)P(B)$.

In the context of machine learning, the MI between features in a dataset can be measured by partitioning the features into two groups, assigning the collective state of each group to variables $A$ and $B$ respectively, and then measuring the amount of correlation that exists between the partitions. This parallels the bipartitioning of the quantum many-body system discussed in Sec.~\ref{sec correlation-scaling-qm}, and allows us to explore \textit{MI scaling} in a similar manner to entanglement scaling. 

\subsection{Entanglement as a Bound on Mutual Information for Orthogonal Data}\label{sec correlation-scaling-bound}

Given the connection between entanglement and tensor networks discussed in Sec.~\ref{sec correlation-scaling-qm}, and having introduced the MI as a classical measure of correlation in Sec.~\ref{sec correlation-scaling-mi}, we now show how the correlations in a classical dataset can guide the choice of network for machine learning. We focus on probabilistic classification, where the tensor network is used to approximate a probability distribution $P(X)$ of feature tensors generated from a classical data distribution $P(\vec{x})$ via Eq.~\eqref{eq prod_state_feat}. We show that for orthonormal inputs the entanglement of the tensor network between feature partitions $A$ and $B$ provides an upper bound on the MI of $P(X)$ between those same partitions. When designing a tensor network for a machine learning task, this relationship can be inverted so that the known MI of a given $P(X)$ sets a lower bound on the entanglement needed for the network to represent it. For non-orthogonal inputs these bounds do not hold rigorously, but may still serve as a useful heuristic for samples with negligible overlap. 

To begin, let $P(\vec{x})$ be the probability distribution associated with feature vectors $\vec{x}$ of length $d$ corresponding to some set $\mathcal{F}$ of $d$ features. Using a tensor-product map of the form in Eq.~\eqref{eq prod_state_feat}, we can map the set of feature vectors $\{\vec{x}\}$ to a set $\mathcal{X}$ of orthogonal rank-one tensors $X \in \mathcal{X}$, generating a new distribution $P(X)$ from $P(\vec{x})$. The overlap of two tensors $X^{(i)}$ and $X^{(j)}$ is determined by the scalar products of the local feature maps
\begin{equation}\label{eq tensor-overlap}
    \braket{X^{(i)}, X^{(j)}} = \prod^d_{k=1}\braket{f_k(x^{(i)}_k), f_k(x^{(j)}_k)} \enspace = \enspace 
    \tikz[baseline=1ex, data/.style={draw, fill=gray!40, minimum size=7}] {
        \node[data] (1_down) {};
        \node[data] (1_up) [above=0.25 of 1_down] {};
        \draw[black, thick] (1_down.north){} -- (1_up.south);
        \foreach \x [count=\i] in {2,3} {
            \node[data] (\x_down) [right=0.25 of \i_down] {};
            \node[data] (\x_up) [above=0.25 of \x_down] {};
            \draw[black, thick] (\x_up.south) -- (\x_down.north);
        }
}\enspace ... \enspace \tikz[baseline=1ex, data/.style={draw, fill=gray!40, minimum size=7}] {
        \node[data] (4_down) {};
        \node[data] (4_up) [above=0.25 of 4_down] {};
        \draw[black, thick] (4_up.south) -- (4_down.north);
    } \enspace ,
\end{equation}
where each feature map is a function of only a single feature. For this analysis we require that the vectors in the image of each local feature map must form an orthonormal set, so that a pair of feature vectors $\vec{x}^{(i)}$ and $\vec{x}^{(j)}$ will always be mapped to either the same tensor or to a pair of orthogonal tensors. For continuous features, such a mapping can be achieved by discretizing the real numbers into $b$ bins, and then assigning values in each bin to a different $b$-dimensional basis vector. The $f_i$ for this mapping will never be one-to-one, although as the dimensionality of their outputs grows the functions will come closer to being injective in practice.

Assuming that the images of the local feature maps are finite-dimensional, $\mathcal{X}$ will be finite and therefore $P(X)$ will be a discrete distribution that can be represented as a tensor $W$ of the form
\begin{equation}\label{eq15}
    W = \sum_{X \in \mathcal{X}} \sqrt{P(X)}X,
\end{equation}
where we have taken the square-root to ensure that $W$ is normalized (i.e. $\braket{W,W} = 1)$.
With this representation, the probability of a given tensor $X$ can be extracted by taking the square of its scalar product with $W$
\begin{equation}\label{eq16}
    P(X) = |\braket{X, W}|^2.
\end{equation}
In the context of machine learning, $W$ can be described using the language of Sec.~\ref{sec tensor-network-ml} as an idealized weight tensor which we seek to model using a tensor network. For a given network, we want to know which $W$, and therefore which $P(X)$, can be accurately represented.

To probe the correlations within $P(X)$, we partition the features into disjoint sets $\mathcal{A}$ and $\mathcal{B}$ such that $\mathcal{A} \cap \mathcal{B} = \emptyset$ and $\mathcal{A} \cup \mathcal{B} = \mathcal{F}$. Using this grouping, the underlying feature distribution $P(\vec{x})$ can be represented as the joint distribution $P(\vec{x}_A, \vec{x}_B)$, where $\vec{x}_A$ and $\vec{x}_B$ are vectors containing values for the features in partitions $\mathcal{A}$ and $\mathcal{B}$ respectively. Similarly, $P(X)$ can be represented as the joint distribution $P(X_A, X_B)$, where $\mathcal{X_A} \ni X_A$ and  $\mathcal{X_B} \ni X_B$ are sets of orthogonal tensors created from the local maps of features in $\mathcal{A}$ and $\mathcal{B}$ respectively. For any tensor $X \in \mathcal{X}$, we have $X = X_A \otimes X_B$ for some $X_A$ and $X_B$. We can also define the marginal distributions $P(X_A)$ and $P(X_B)$ that describe the statistics within each partition separately. The MI $I(X_A : X_B)$ across the bipartition is given as in Eq.~\eqref{eq13} using the entropies of these distributions.

To introduce the entanglement measure described in Sec.~\ref{sec correlation-scaling-qm} as a bound on $I(X_A : X_B)$, we represent the normalized tensor $W$ as the quantum state $\ket{\psi_W}$ and the tensors in $\mathcal{X}$ as orthonormal basis states $\ket{X_A,X_B}$, such that Eq.~\eqref{eq15} becomes
\begin{equation}
    \ket{\psi_W} = \sum_{\mathcal{X_A},\mathcal{X_B}}\sqrt{P(X_A, X_B)}\ket{X_A,X_B},
\end{equation}
where we have shifted to ket notation. This encoding of a probability distribution into a quantum state has been utilized previously in the study of quantum Bayesian algorithms~\cite{Low_Yoder_Chuang_2014}.
The process of extracting $P(X_A,X_B)$ described in Eq.~\eqref{eq16} can be reimagined as projective measurements of $\ket{\psi_W}$ on an orthonormal basis, where the probabilities are used to reconstruct $P(X_A,X_B)$. Since the MI between outcomes of local measurements on a quantum state is upper bounded by the entanglement of that state~\cite{Wu_Poulsen_Molmer_2009}, $\ket{\psi_W}$ must have a bipartite entanglement with respect to partitions $\mathcal{A},\mathcal{B}$ that is at least as large as $I(X_A: X_B)$. The MI of $P(X)$ across a bipartition therefore provides a lower bound on the amount of entanglement needed in $\ket{\psi_W}$ with respect to that same partition 
\begin{equation}\label{eq mi_ent_bound}
    I(X_A: X_B) \leq E(\mathcal{A},  \mathcal{B}),
\end{equation}
which through Eq.~\eqref{eq network entangle} sets a lower bound on the degree of connectivity $n$ and/or bond dimension $m$ needed in the tensor network representing $\ket{\psi_W}$.

In a typical machine learning setting, we will have access to samples of $P(\vec{x})$, which can then be encoded into tensors which form samples of $P(X)$. If we aim to estimate the MI numerically, as we will in Sec.~\ref{sec mi-estimation}--\ref{sec real-data}, then it is generally easier to work with the original feature vectors sampled from $P(\vec{x})$ than with the feature tensors from $P(X)$. From the data processing inequality~\cite{Cover_Thomas_1991}, $I(X_A,X_B)$ is upper-bounded by $I(\vec{x}_A, \vec{x}_B)$, so using the MI of the original features will yield a bound on the entanglement that may be larger than necessary to model $P(X)$, but will always be sufficient. Indeed, as the dimensionality of the feature map outputs increases, the gap between $I(X_A,X_B)$ and $I(\vec{x}_A, \vec{x}_B)$ will shrink---since the finer discretization preserves more information---and thus the estimates from both featurizations will converge.

The methodology described above may appear somewhat circuitous, in that we start from the tensorized entanglement formalism that is most natural for tensor networks, but then move back to a classical MI description of the original data features. At first glance it seems like a more direct approach would be to simply estimate the entanglement of $\ket{\psi_W}$ between partitions $A$ and $B$ directly, using some approximation $\ket{\Tilde{\psi}_W}$ constructed from the available data
\begin{equation}
    \ket{\Tilde{\psi}_W} \propto \sum^N_{i=1} \ket{X^{(i)}_A, X^{(i)}_B},
\end{equation}
where $\{\ket{X^{(i)}_A, X^{(i)}_B}\}$ is a set of $N$ of samples from $P(X_A, X_B)$. Such a construction was recently used for entanglement analysis by Martyn et al.~\cite{Martyn_Vidal_Roberts_Leichenauer_2020} in the context of MPS image classification. Unfortunately, as evident in \cite{Martyn_Vidal_Roberts_Leichenauer_2020}, the entanglement of $\ket{\Tilde{\psi}_W}$ is artificially upper-bounded by $\log(N)$, independent of the actual properties of $P(X_A,X_B)$. This saturation occurs because, for generic sample tensors $\ket{X^{(i)}}$ and $\ket{X^{(j)}}$ with $d$ features, we have
\begin{equation}\label{eq tensor_overlap_supress}
\braket{X^{(i)}|X^{(j)}} = \prod^d_{k=1} \braket{f_k(x^{(i)}_k), f_k(x^{(j)}_k)} \approx c^d
\end{equation}
for some typical local overlap $c < 1$. As the number of features grows, the overlap between data tensors is exponentially suppressed. When calculating the entanglement, the near-orthogonality of tensors within $\mathcal{X}_\mathcal{A}$ and  $\mathcal{X}_\mathcal{B}$ (when partitions $A$ and $B$ are both moderately sized) causes the partial trace to generate an almost maximally mixed state with a von Neumann entropy of approximately $\log(N)$. In contrast, by moving back to the original vector space of the data and using MI rather than entanglement, we can generally avoid the $\log(N)$ upper bound (in Sec.~\ref{sec discussion} we discuss specific circumstances where this limit can also appear in MI estimation).

\section{Estimating Mutual Information}\label{sec mi-estimation}

\subsection{Setup and Prior Work}

For our analysis in Sec.~\ref{sec correlation-scaling} to be of practical use, we need a method of estimating the MI of a probability distribution using only a finite number of samples. More formally, let $\{\vec{x}^{(i)}\}$ be a set of $N$ samples drawn from a distribution $P(\vec{x})$ whose functional form we do not, in general, have access to. For a bipartition $\mathcal{A},\mathcal{B}$ of the dataset features, our goal is to estimate the MI of $P(\vec{x}_A, \vec{x}_B)$ between the features in $\mathcal{A}$ and the features in $\mathcal{B}$ using these samples. 

Several approaches to MI estimation~\cite{Paninski_2003} have been proposed and explored in the literature. For continuous variables, some methods discretize the variable space into bins, and then compute a discrete entropy value based on the fraction of samples in each bin~\cite{Moddemeijer_1989}\cite{Steuer_Kurths_Daub_Weise_Selbig_2002}. Alternatively, kernel density estimators~\cite{Epanechnikov_1969} can be used to directly approximate the continuous probability density function using a normalized sum of window functions centered on each sample, which is then used to calculate the MI~\cite{Moon_Rajagopalan_Lall_1995}. A method developed by Kraskov et al.~\cite{Kraskov_Stogbauer_Grassberger_2004}, which utilizes a \textit{k}-nearest neighbor algorithm to calculate the MI, has become popular due to its improved error cancellation when calculating the MI from approximated entropies.

For this paper, we base our estimation method on more recent work by Koeman and Heskes~\cite{Koeman_Heskes_2014} and Belghazi et al.~\cite{Belghazi_Baratin_Rajeswar_Ozair_Bengio_Courville_Hjelm_2018}. In \cite{Koeman_Heskes_2014}, the MI estimation problem is recast as a binary classification task between samples from $P(\vec{x}_A, \vec{x}_B)$ and $P(\vec{x}_A)P(\vec{x}_B)$, which the authors modeled using a random forest algorithm. In \cite{Belghazi_Baratin_Rajeswar_Ozair_Bengio_Courville_Hjelm_2018}, Belghazi et al. use a neural network to perform unconstrained optimization on the \textit{Donsker-Varadhan representation} (DV-representation)  of the KL-divergence between $P(\vec{x}_A, \vec{x}_B)$ and $P(\vec{x}_A)P(\vec{x}_B)$, which provides a lower-bound on the MI. In our work, we found that a mixture of these two approaches was most effective. Specifically, we have used the binary classification framing proposed in \cite{Koeman_Heskes_2014}, but approached the problem as a logistic regression task optimized using maximum log-likelihood on a neural network. To evaluate the MI, we used the DV-representation as in \cite{Belghazi_Baratin_Rajeswar_Ozair_Bengio_Courville_Hjelm_2018} to generate a lower-bound when possible. In practice this also gave us smoother MI curves and smaller errors. To our knowledge this overall approach has not be reported in the literature, though it appears similar in concept to a method proposed by Pool et al.~\cite{Poole_Ozair_Oord_Alemi_Tucker_2019} in the context of generative adversarial networks. In the next subsection we describe our algorithm in more detail.     

\subsection{Logistic Regression for MI Estimation}

The logistic regression approach to MI estimation is built around the KL-divergence definition of the MI introduced in Eq.~\eqref{eq mi-kl}. In the context of our dataset, the variable spaces $\mathcal{A}$ and $\mathcal{B}$ describe the collective values of the features in partitions $\mathcal{A}$ and $\mathcal{B}$ respectively, with the sums taken over all allowed value combinations. For convenience, we simplify our notation such that $a \equiv \vec{x}_A$ and $b \equiv \vec{x}_B$ represent the feature values of each partition. To estimate the MI using the KL-divergence, we require an approximation for $f(a, b) = \log\frac{P(a, b)}{P(a)P(b)}$. This can be found via logistic regression by first recasting the joint and marginal probability distributions as conditional probabilities
\begin{equation}\label{eq prob_def}
    P(a,b|\text{joint}) \equiv P(a,b), \quad \quad
    P(a,b|\text{marg}) \equiv P(a)P(b),
\end{equation}
where $P(a, b|\text{joint})$ is the probability that the feature values $a, b$ will be sampled from the joint distribution $P(a, b)$, and $P(a, b|\text{marg})$ is the probability that the values will be sampled from the product-of-marginals distribution $P(a)P(b)$. Using Bayes' theorem, the conditional probabilities can be reversed
\begin{equation}\label{eq bayes_theorem}
    P(a, b|\text{joint}) \propto \frac{P(\text{joint}|a, b)}{P(\text{joint})}, \quad \quad P(a, b|\text{marg}) \propto \frac{P(\text{marg}|a, b)}{P(\text{marg})}.
\end{equation}
Substituting Eq.~\eqref{eq prob_def} and Eq.~\eqref{eq bayes_theorem} back into $\log\frac{P(a, b)}{P(a)P(b)}$ gives
\begin{equation}\label{eq:log_ratio}
    \log\frac{P(a, b)}{P(a)P(b)} = \log\frac{P(\text{joint}|a, b)}{P(\text{marg}|a, b)} + \log\frac{P(\text{marg})}{P(\text{joint})},
\end{equation}
where the first term is the log-odds of a binary classification problem where samples are taken from either $P(a, b)$ or $P(a)P(b)$ and the classifier must decide the most likely source for a given set of feature values $a$ and $b$. The second term will equal zero if each source is equally likely to be sampled.

To get a numerical estimate of Eq.~\eqref{eq:log_ratio}, we can train a parameterized function $T(a, b)$ to estimate the log-odds via standard logistic regression methods
\begin{equation}\label{eq22}
    T(a, b) \approx \log\frac{P(\text{joint}|a, b)}{P(\text{marg}|a, b)},    
\end{equation} 
using a training set that consists of an equal number of joint samples and marginal samples. In particular, we parameterized $T$ using a dense feed-forward neural network to avoid introducing spatial bias, and optimized the network by minimizing the binary cross-entropy (i.e. maximizing the log-likelihood) across the samples.

Since the joint distribution is the actual source of our dataset, we already have $N$ samples from it. To approximate a sample from the product-of-marginals distribution, we take a set of values for the features in  $\mathcal{A}$ from a joint sample chosen at random, and then take values for the features in $\mathcal{B}$ from another randomly-chosen joint sample (the two sources could be the same sample, although this is unlikely for a large dataset). After selection, the features are combined together into a single mixed sample which, by construction, has no correlations across the partition. After training the network, the MI could be estimated by taking the average of $T$ across the joint samples as a direct approximation\footnote{For $M$ samples of $P(a, b)$, we have $\lim\limits_{M \rightarrow \infty}\frac{1}{M}\sum\limits^M_{i=1} T(a_i, b_i) = \sum\limits_{a \in \mathcal{A}, b \in \mathcal{B}} P(a, b)T(a, b)$.} of the KL-divergence from Eq.~\eqref{eq mi-kl}
\begin{equation}\label{eq direct-est}
    I(A: B) \approx \frac{1}{M}\sum^M_{i=1}T(a_i, b_i),
\end{equation}
where $a_i$ and $b_i$ are the feature values of the $i$th joint sample taken from a validation set of size $M$. However, a superior approach is to insert $T$ into the DV-representation~\cite{Ruderman_Reid_Garcia-Garcia_Petterson_2012} of the MI
\begin{equation}\label{eq dv-rep}
I(A : B) \geq \frac{1}{M}\sum^M_{i=1}T(a_i, b_i)-\log\frac{1}{M^2}\sum^M_{i=1}\sum^M_{j=1}e^{T(a_i, b_j)},
\end{equation}
which yields a lower-bound on the MI as $M \rightarrow \infty$ and allows errors to cancel\footnote{
For example, errors of the form $T(a, b) + \epsilon$ can be brought outside of the sums in Eq.~\eqref{eq direct-est} to give $\epsilon - \log e^\epsilon$, which cancels. 
}. The inequality is saturated when $T = \log\frac{P(a, b)}{P(a)P(b)}$, since as $M \rightarrow \infty$ the second term vanishes and the first term gives the KL-divergence. Belghazi et al. carried out their MI estimation by maximizing Eq.~\eqref{eq dv-rep} itself, but we have found in practice that the second term often overflows on datasets with large MI. Furthermore, the optimization algorithm would often attempt to maximize the second term, even though it must vanish in the optimal solution. We were able to mitigate these problems by instead training with the binary cross-entropy~\cite{Ramos_Franco-Pedroso_Lozano-Diez_Gonzalez-Rodriguez_2018} as a loss function and only using Eq.~\eqref{eq dv-rep} at the end to get the MI value of the optimized distribution. As a caveat, we found in practice that for certain distributions with larger MI values Eq.~\eqref{eq direct-est} generally yielded more stable and accurate estimates than Eq.~\eqref{eq dv-rep}, though the reason for this is not clear.

\section{Numerical Tests with Gaussian Fields}\label{sec GMRF}

\subsection{Gaussian Markov Random Fields}

To test the accuracy of the logistic regression algorithm, we need a distribution to sample from that has an analytic expression for the MI and that can model different MI scaling patterns. Both of these requirements are satisfied by Gaussian Markov random fields (GMRFs)~\cite{Rue_Held_Held_2005}, which are multivariate Gaussian distributions parameterized by the \textit{precision matrix} $Q \equiv \Sigma^{-1}$, where $\Sigma$ is the more familiar covariance matrix. With respect to $Q$, the Gaussian distribution with mean $\vec{\mu}$ is
\begin{equation}
    p(\vec{x}) = \sqrt{\text{det}(\frac{1}{2\pi}Q})\exp[-\frac{1}{2}(\vec{x} - \vec{\mu})^TQ(\vec{x} - \vec{\mu})],
\end{equation}
where $p(\vec{x})$ is the probability density of the variables $\vec{x}$. The element $Q_{ij}$ of the precision matrix determines the \textit{conditional correlation} between variables $x_i$ and $x_j$, which describes the statistical dependence of the pair when all other variables are held fixed at known values. This is in contrast with the more familiar \textit{marginal correlation}, governed by $\Sigma$, which describes the dependence between a pair of variables when the state of all other variables is unknown. If $Q_{ij} = 0$, the variables $x_i$ and $x_j$ are conditionally uncorrelated:
\begin{equation}
    p(x_i,x_j|x_{k\neq i,j}) = p(x_i|x_{k\neq i,j})p(x_j|x_{k\neq i,j}) \Longleftrightarrow Q_{ij} = 0.
\end{equation}
By setting specific elements of the precision matrix to zero, the correlation structure and therefore the MI of the Gaussian can be tuned to a desired pattern. This flexibility allows us to encode different MI scaling patterns into the distribution, which can then be extracted analytically using Eq.~\eqref{eq13} and the expression for Gaussian entropy
\begin{equation}
    S = \frac{1}{2}\log[\det(2\pi e \Sigma)],
\end{equation}
which combine together to give an expression for the Gaussian MI:
\begin{equation}\label{eq gauss_mi}
    I(A : B) = S(A) + S(B) - S(AB) =  \frac{1}{2}\log\frac{\det(\Sigma_A)\det(\Sigma_B)}{\det(\Sigma)},
\end{equation}
where $\Sigma_A$ and $\Sigma_B$ are the covariance matrices corresponding to variables in partitions $A$ and $B$ respectively.

\subsection{Test Setup}

In the following subsections, we present test results of the logistic regression estimator on GMRFs representing three different correlation patterns: a boundary law with nearest-neighbor correlations, a volume law with weak correlations across all variables, and a distribution with sparse, randomized correlations. In the language of quantum many-body physics, the first two patterns reflect correlation structures that would be expected in ground states and random states respectively, while the GMRF with random sparse correlations shows the scaling for a heterogeneous distribution that lacks any spatial structure. These Gaussian distributions serve as both a means of testing the algorithm and as a clear illustration of the numerical MI plots that would be expected from different types of correlations within a dataset.

In the tests, each GMRF consisted of 784 variables, which mirrored the number of pixels in the $28 \times 28$ images taken from the MNIST and Tiny Images datasets analyzed in Sec.~\ref{sec real-data}. To measure the scaling behavior of the MI in these GMRFs, we used a range of different bipartition sizes, with the partitions being selected such that they always formed a pair of contiguous patches when the variables were arranged in a $28 \times 28$ array. One member of each bipartition was formed from an inner square patch of variables centered on the array, whose side length we denote as $L$. The other partition was an outer patch consisting of all other variables. The size of the inner partition ranged from a single variable ($L = 1)$ to a $26 \times 26$ block ($L = 26$). For each bipartition, the MI was estimated using our logistic regression algorithm and the DV-representation, with the estimates plotted alongside the analytic MI curve of the GMRF to evaluate their quantitative and qualitative accuracy. Since our model used a stochastic gradient descent method for optimization, we averaged over multiple training runs to generate a representative curve. To explore the effect of sample size on the algorithm, we generated datasets from the GMRFs with 70,000, 700,000, and 7,000,000 joint training samples and created MI curves for each size using averages over 20, 10, and 5 trials respectively. Samples and covariance plots from the GMRF test distributions are given in Sec.~\ref{sec appendix_gmrf}. 

\subsection{Nearest-Neighbor Boundary-Law GMRF}\label{sec nearest-neighbor}

 As shown in Eq.~\eqref{eq boundary law}, for the MI of a bipartition to obey a boundary law its magnitude must scale with the length of the boundary or interface between the partitions. Given a set of variables on a $d$-dimensional lattice, the simplest way to construct a boundary law is to have each variable be conditionally-correlated with only its $2d$ nearest neighbors. For variable $x_{ij}$ on a two-dimensional grid at row $i$ and column $j$, the conditional probability function would depend on the values of only four other variables
 \begin{equation}\label{eq28}
     p(x_{i,j}|\{x_{k \neq i,j}\}) = p(x_{i, j}|x_{i + 1,j}, x_{i-1,j}, x_{i, j+1}, x_{i, j-1}),
 \end{equation}
although the number of neighbors can be fewer if the variable is at an edge or corner since the grid is finite. After partitioning, the inner patch of variables will be conditionally correlated with only a single layer of variables surrounding its perimeter, so the MI between the inner and outer partitions will be proportional to $L$. To encode the correlation structure of Eq.~\eqref{eq28} into a precision matrix, all of the off-diagonal elements in each row of $Q$ must be set to zero except those that correspond to the nearest neighbors, with the non-zero off-diagonal elements all assigned the same value $q$ that determines the strength of the correlation. To guarantee that $Q$ is positive definite, $q$ should not exceed the magnitude of the diagonal elements divided by the number of nearest neighbors (see Sec.~\ref{sec uniform-law} for more details on these constraints).

\begin{figure}
    \centering
    \includegraphics[width=\textwidth]{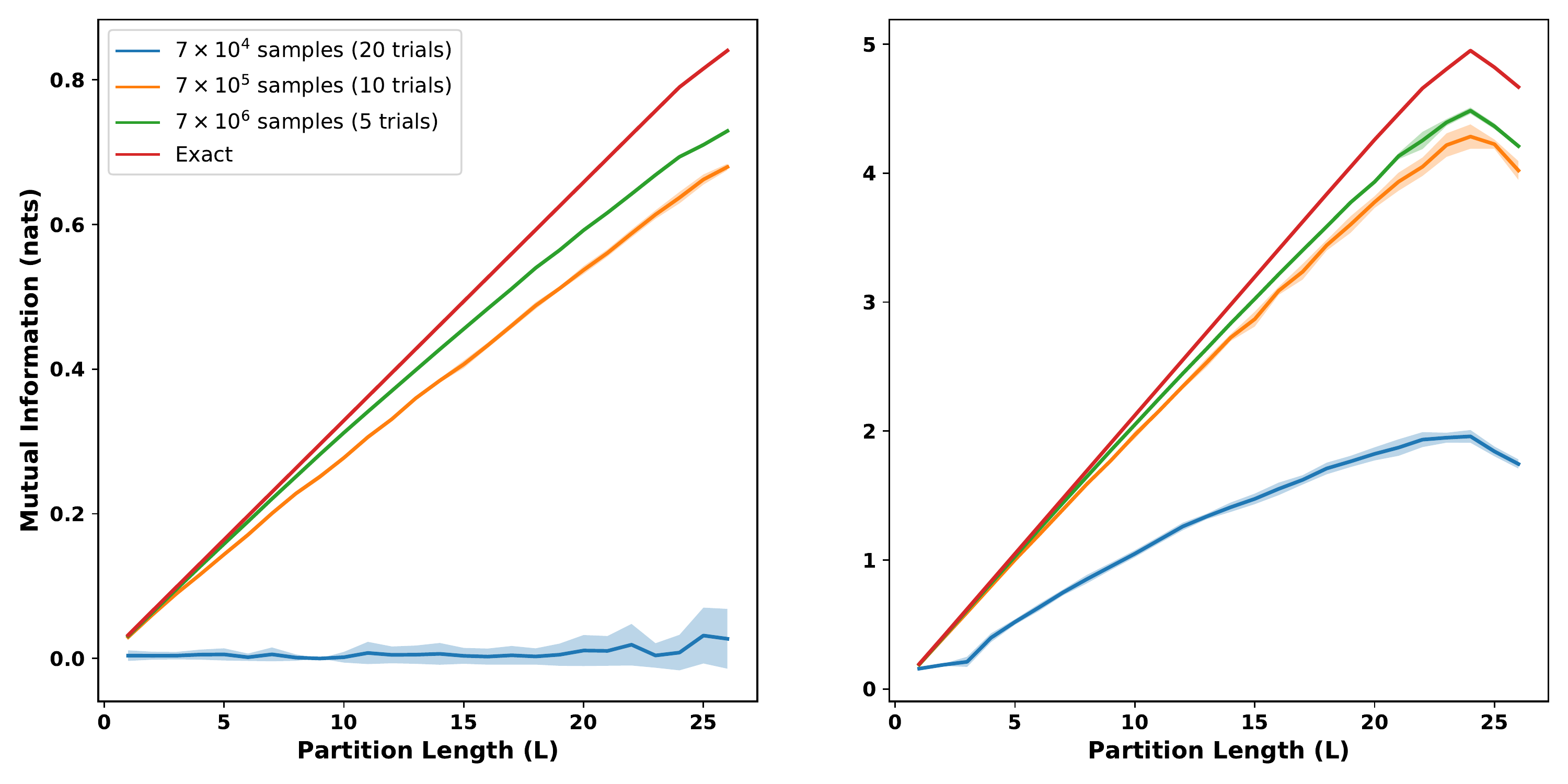}
    \caption{MI curves for a GMRF with nearest-neighbor correlations at different sample sizes, plotted relative to the side length $L$ of the inner partition. The plot on the left had a weaker correlation strength with $q = -0.12$, while the plot on the right had a stronger correlation strength with $q = -0.227$. The solid lines represent the averages over the trials, while the shaded regions show one standard deviation. The linear boundary-law scaling pattern of the GMRF is evident from the exact curve (red). With the exception of the weakly-correlated, 70,000 sample trial, this linear scaling is successfully reconstructed by the algorithm. The magnitude of the MI is underestimated in all trials, with the fractional error being similar for the strong and weak correlations.}
    \label{area_large_small}
\end{figure}

The performance of the logistic regression algorithm on the nearest-neighbor GMRF is summarized in Figure \ref{area_large_small}, which plots the MI in \textit{nats} against the side length $L$ of the square inner partition\footnote{When calculating quantities such as the entropy or MI using natural logarithms, the unit of information is a \textit{nat} instead of a \textit{bit}.
}. Since the x-axis is proportional to the perimeter of the inner patch rather than its area, we expect a boundary-law MI curve to be linear in $L$. This is clearly evident in the analytic curve, which is linear up to a length of roughly 25 variables before leveling off. The linear pattern is broken near the boundaries because the marginal correlations between variables around the edges of the grid are smaller than those between variables closer to the center. Aside from the the 70,000 sample trial with weak correlations, the regression estimates were able to successfully reproduce the boundary-law scaling pattern, with the error shrinking as the number of samples increased. It is also interesting to note that the fractional errors of the different sample sizes are similar between the strong and weak correlations, suggesting that the source of the error is independent of the MI magnitude. 

\subsection{Uniform Volume Law GMRF}\label{sec uniform-law}

In contrast with the local correlations that give rise to a boundary law, we can imagine an alternative pattern in which each variable is equally correlated with every other variable. These correlations produce a volume law for the MI, since every variable in the inner partition must contribute equally to the correlations with the outer partition. To encode such a pattern into a GMRF, we set every off-diagonal element of the precision matrix $Q$ to the same value $q$. To ensure that the precision matrix remains positive definite, the value $q$ should be small enough to preserve \textit{diagonal dominance}, a sufficient but not strictly necessary condition for a positive definite matrix in which the sum of the magnitudes of the off-diagonal elements of a row or column do not exceed the diagonal element
\begin{equation}\label{eq diag-dom}
    Q_{ii} > \sum_{i \neq j}|Q_{ij}| \text{\  and \ } Q_{ii} > \sum_{j \neq i} |Q_{ij}|.
\end{equation}
To create a uniform scaling pattern it suffices to set $Q_{ii} = 1$, which means we must have $q < \frac{1}{N-1}$ for an $N$-dimensional Gaussian. This provides an upper limit on the amount of correlation one Gaussian variable and can have with any other when the correlations are homogeneous, a limit that decreases as the number of variables grows larger.

\begin{figure}
    \centering
    \includegraphics[width=\textwidth]{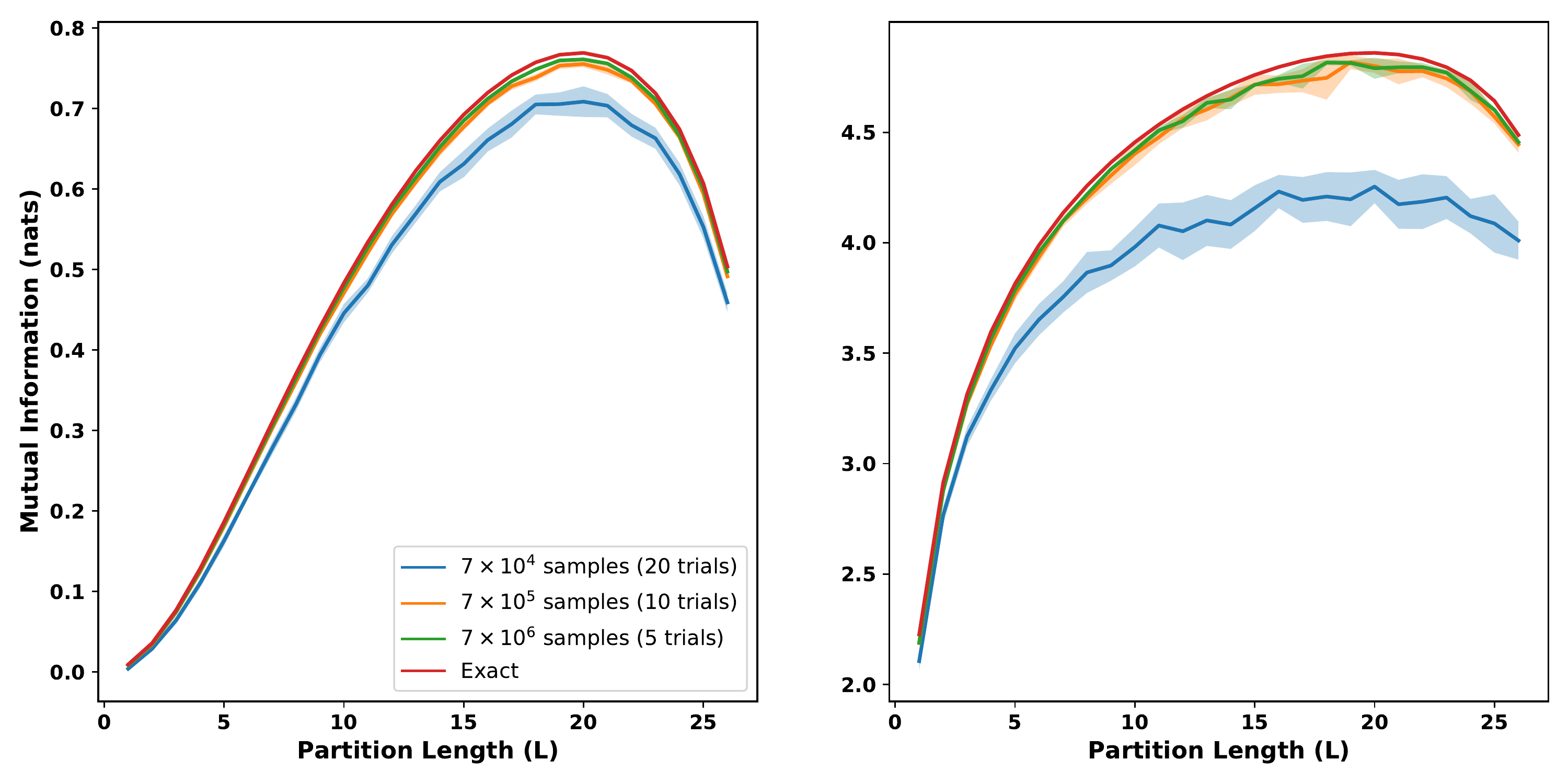}
    \caption{MI curves for a GMRF with uniform correlations at different sample sizes, plotted relative to the side length of the inner partition. The plot on the left had a weaker correlation strength with $q = -1.2\times10^{-3}$, while the plot on the right had a stronger correlation strength with $q = -1.27712\times10^{-3} \approx \frac{1}{783}$. The solid lines represent the averages over the trials, while the shaded regions show one standard deviation. The algorithm successfully reproduced the shape of the exact MI curve, with the larger sample sizes almost matching the analytic MI values. The finite size of the grid causes the curve to gradually bend over as the partition length increases.}
    \label{fig:diffuse_large_small}
\end{figure}

The performance of our algorithm on a GMRF with these uniform correlations is summarized in Figure \ref{fig:diffuse_large_small}. We were able to accurately reproduce the shape and approximate magnitude of the analytic curves for both correlation strengths and for all sample sizes, although as expected the 70,000 sample trials had the largest error. Interestingly, the algorithm performed significantly better on the uniform GMRF than on the nearest-neighbor GMRF, even though the pairwise dependence between correlated variables in the former was much weaker than in the latter. This suggests that, for a given amount of MI, it is easier for the algorithm to find correlations that are spread out across many variables than to identify those that are concentrated in some sparse set.

It is worth noting that the shape of a volume-law curve should be quadratic on the axes used in Figure \ref{fig:diffuse_large_small}, yet from our plots it is clear that the quadratic form breaks down quickly for the weak correlations and never exists at all for the strong correlations. This distortion occurs because the MI is purely a function of the number of variables in each partition when the correlations are homogeneous, and due to the finite size of our grid any increase in the size of the inner patch necessarily comes at the cost of the outer patch. Correspondingly, any increase in the MI that comes from growing the inner patch is partially offset by the correlations that are lost when shrinking the outer patch. On the $28 \times 28$ grid used in Figure \ref{fig:diffuse_large_small}, the MI begins to decline at partition length $L = 20$, which marks the point where both partitions contain roughly the same number of variables (400 vs 384) and where the amount of correlation is therefore maximized.

\subsection{Random Sparse GMRF}

A third class of GMRF to explore is one where the correlations have no inherent spatial pattern yet are also non-uniform. Such a distribution could, for example, represent a dataset of features that are correlated but lack the in-built sense of position or ordering necessary to unambiguously map them onto a lattice (e.g., demographic data). If we nevertheless insist on embedding these features into a grid, we can expect that for most arrangements the MI will scale either as a volume law or in some irregular pattern, depending on whether the features all have similar correlation strengths.

For our tests, we engineered a spatially-disordered GMRF by taking the nearest-neighbor precision matrix used in Sec.~\ref{sec nearest-neighbor} and randomly permuting the variables around the grid. Under this scheme, each row and column of the precision matrix $Q$ has four non-zero off-diagonal elements in random positions. While the conditional correlations of this new distribution are still sparse, they are no longer exclusively short-range but can instead span the entire grid. Since all of the non-zero off-diagonal elements of $Q$ have the same magnitude $q$, the amount of correlation across any bipartition increases evenly with the number of correlated variable pairs shared between the partitions. Without any underlying spatial structure, the odds of a given pair being separated into two different partitions is roughly proportional to the volume of the smaller partition, assuming that the other partition is much larger. Under the inner-outer partitioning scheme used in our tests, we expect a volume law for small partition lengths, followed by the same bending-over observed in Sec.~\ref{sec uniform-law} for the uniform correlations. 

\begin{figure}
    \centering
    \includegraphics[width=\textwidth]{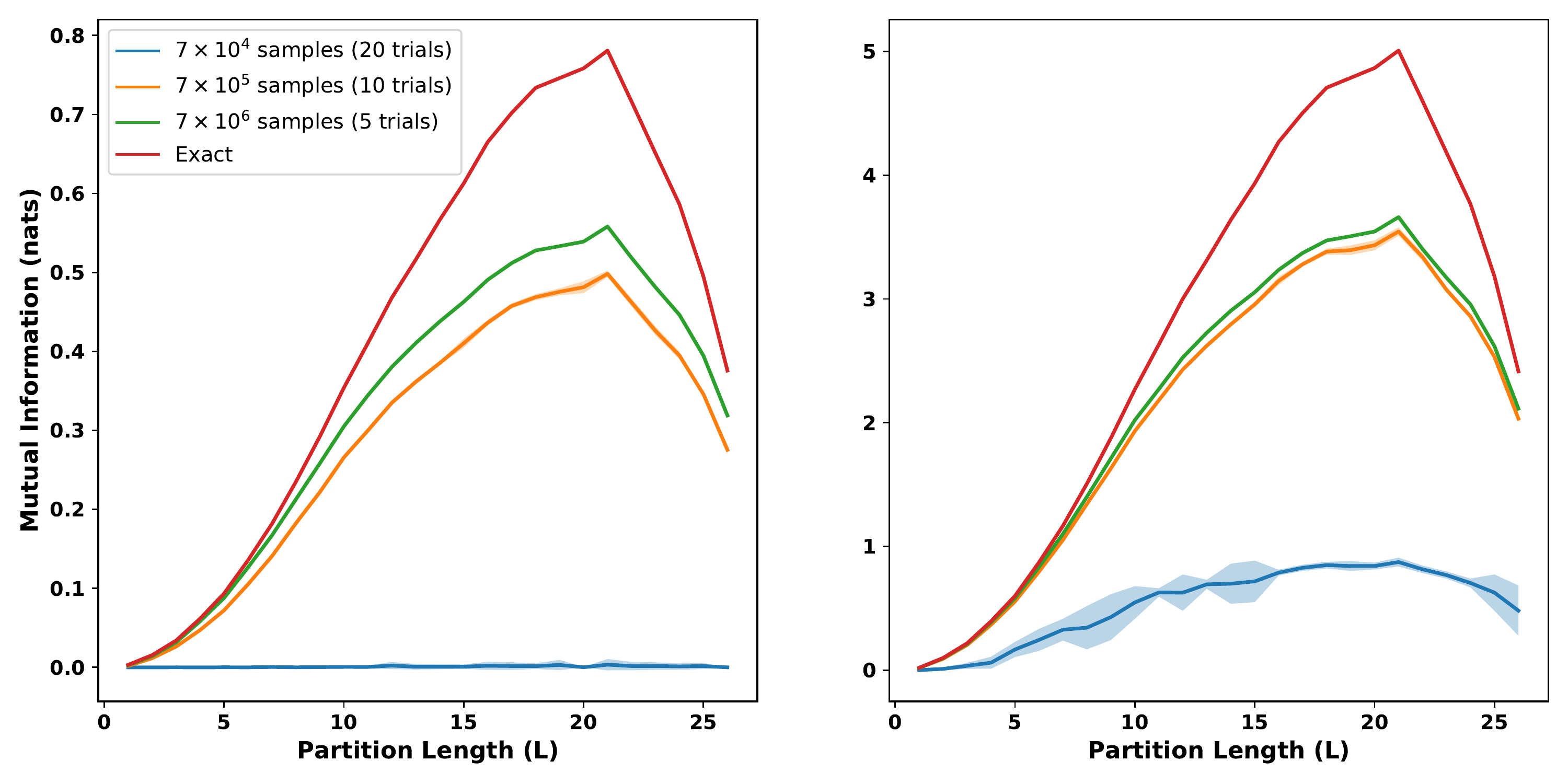}
    \caption{MI curves for a GMRF with spatially-randomized correlations at different sample sizes, plotted relative to the side length of the inner partition. The plot on the left had a weaker correlation strength with $q = -0.045$, while the plot on the right had a stronger correlation strength with $q = -0.11$. The solid lines represent the averages over the trials, while the shaded regions show one standard deviation. The quadratic, volume-law scaling is clear from the analytic MI curve at small $L$, which for the strong correlations was reproduced across all sample sizes (though the 70,000 sample curve is greatly diminished). For the weak correlations the model was unable to find any correlations using the smallest sample size of 70,000, as was the case for the nearest-neighbor correlations (Figure \ref{area_large_small}, left panel).}
    \label{sparse_large_small}
\end{figure}

The performance of our logistic regression algorithm on a GMRF with these spatially-randomized correlations is shown in Figure \ref{sparse_large_small}. As predicted, the analytic curves show similar scaling patterns to those of the uniform GMRF in Figure~\ref{fig:diffuse_large_small}. The quality of the MI estimates, however, is more similar to the nearest-neighbor MI curves of Figure \ref{area_large_small}, where the model succeeded at replicating the analytic MI curve for all sample sizes when the correlation strength was large, but failed for the smallest sample size (70,000) when the correlations were weak. The estimation error is larger overall for the randomized variables than for the nearest-neighbor variables, and increasing the sample sizes appears to yield diminishing returns. This may partially stem from the reduction in pairwise correlation strength (quantified by $q$) that was required to keep the magnitude of the peak MI value consistent between the different GMRFs. However, a more likely explanation is that the nearest-neighbor correlations are able to reinforce one another due to their shared proximity, which results in the next-nearest-neighbor marginal correlations also being quite strong. This may make the correlations easier for a machine learning algorithm to detect, since they will impact a larger number of variables. In contrast, for the randomized GMRF the correlated variables are scattered far away from each other on average, which severely diminishes any reinforcement effect.

\section{Application to Image Data}\label{sec real-data}

\subsection{Setup}

To explore the types of MI scaling patterns that might be seen in real data, we analyzed two sets of images: the 70,000 image MNIST handwritten-digits dataset~\cite{MNIST}, and 700,000 images taken from the Tiny Images dataset~\cite{Torralba_Fergus_Freeman_2008} converted to grayscale using a weighted luminance coding\footnote{For normalized RGB color values, each grayscale pixel was assigned the value 0.3R + 0.59G + 0.11B. See \cite{Kanan_Cottrell_2012} for more information on grayscale conversions.}. Sample images from these datasets and further details can be found in Sec.~\ref{sec appendix_data}. These two datasets were chosen due to their differing levels of complexity: MNIST consists of simple, high-contrast shapes while the Tiny Images are low-resolution depictions of the real world with much more subtle color gradients. In our experiments, each image contained 784 pixels arranged in a $28 \times 28$ array, with the Tiny Images dataset being cropped from $32 \times 32$ by removing two pixels from each side. The pixel values, originally integers from 0 to 255, were rescaled to the range [0, 1].

To generate the MI estimates for these two datasets, we used the same partitioning method described in Sec.~\ref{sec GMRF} for the GMRFs, with each image being split into a centered, square inner patch of increasing size and a surrounding outer patch. These partitions were then fed into the algorithm laid out in Sec.~\ref{sec mi-estimation}, with one key difference; the DV-representation of Eq.~\eqref{eq dv-rep} proved to be unusable for both MNIST and the Tiny Images due to instability in the exponential term. While we were able to use the DV-representation to significantly reduce error on the GMRF tests, on the real datasets we had to instead make a direct estimate of the KL-divergence from Eq.~\eqref{eq direct-est}. It is not clear why the DV-representation worked for the GMRFs but not for the image datasets, although this could be due to the larger MI and stronger correlations that are present in the real-world data. 

\subsection{Results}

\begin{figure}
    \centering
    \includegraphics[width=0.9\textwidth]{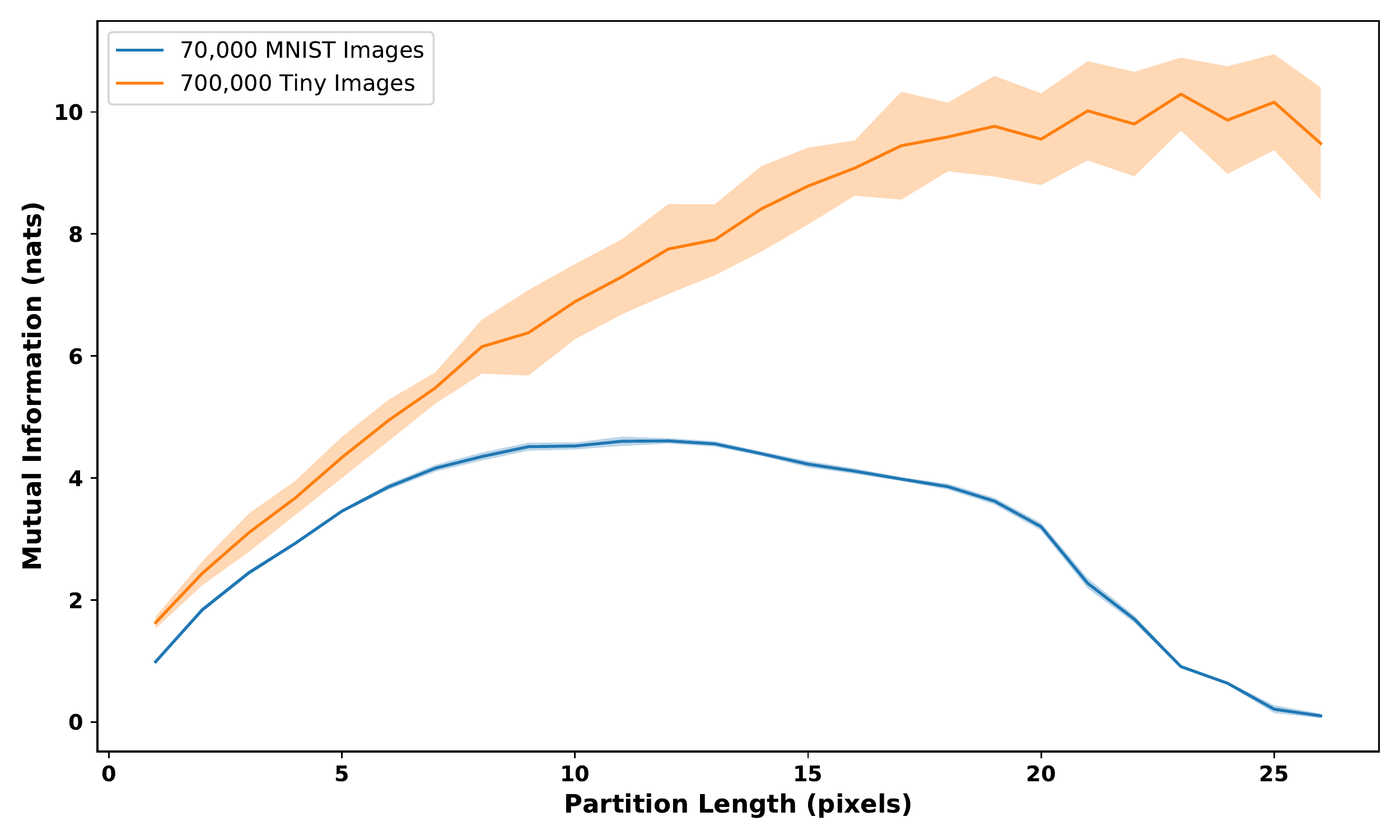}
    \caption{MI estimates for the MNIST and Tiny Images datasets using logistic regression, plotted relative to the side length $L$ of the inner partition. The solid lines are averages from twenty separate trials, while the shaded regions show one standard deviation. The MNIST curve most closely resembles the strongly-correlated uniform GMRF from Sec.~\ref{sec uniform-law}, and exhibits minimal variance. The Tiny Images curve is most similar to the nearest-neighbor, boundary-law GMRF from Sec.~\ref{sec nearest-neighbor} (Figure \ref{area_large_small}, right panel), but the shape is harder to pin down due to its high variance.} 
    \label{mnist_tiny}
\end{figure}

Figure \ref{mnist_tiny} shows the MI of the MNIST and Tiny Images datasets as estimated by logistic regression, plotted relative to the side length $L$ of the inner pixel partition. The MI curves were generated from averages taken over twenty different trials, and plotted within a shaded region containing one standard deviation. As with the GMRFs, this averaging helped smooth the curves and make their shapes easier to assess, especially for patch sizes with larger variance. 

Looking first at the Tiny Images curve, we can see a moderately linear segment from 1 pixel length to roughly 18 pixels length, which then flattens out and begins to decrease at the $26 \times 26$ patch. Of the three scaling curves tested in Sec.~\ref{sec GMRF}, this overall shape is most consistent with the boundary-law scaling pattern of Sec.~\ref{sec nearest-neighbor} (Figure \ref{area_large_small}, right panel). Unfortunately the variance of the algorithm increased significantly at larger MI values, making it more difficult to assess the pattern. For MNIST, the MI curve most closely resembles that of the strongly-correlated uniform GMRF (Figure \ref{fig:diffuse_large_small}), rising at a decreasing rate until it crests and gradually declines. However, this shape is not as distinct as that of a linear or quadratic curve, so it is difficult to use as evidence for a volume law. 

Interestingly, the MNIST curve shows far less variance than the Tiny Images curve, despite the fact that it contains only a tenth of the images. For the GRMF tests done in Sec.~\ref{sec GMRF}, there was a clear reduction in the variance of each curve as the sample size increased, but this not observed in Figure~\ref{mnist_tiny}. Indeed, the MNIST curve has a smaller variance at each patch size than the Tiny Images curve has at almost any patch size, even when the MI of the MNIST curve is larger. This suggests that there is some data-specific effect causing the discrepancy, perhaps attributable to the relative simplicity of the MNIST images relative to the more realistic Tiny Images.  

Unlike in our GMRF tests, we do not have access to the underlying probability distributions that MNIST and the Tiny Images datasets were sampled from, so it is much more difficult to assess the accuracy of the curves in Figure \ref{mnist_tiny}. One approximate way of evaluating the estimates is to fit a GMRF to the empirical covariance matrix of the data, and then calculate the Gaussian MI analytically in the same manner as in Sec.~\ref{sec GMRF}. This new distribution is constrained to model only pairwise interactions between the variables, and all marginal and conditional distributions among the variables are forced to be Gaussian, so it is not representative of the true distribution. Nevertheless, due to its high entropy and simple correlation structure, a fitted GMRF is likely (but not guaranteed~\cite{Pires_Perdigao_2012}) to provide a lower bound on the MI of the true distribution.

\begin{figure}
    \centering
    \includegraphics[width=0.9\textwidth]{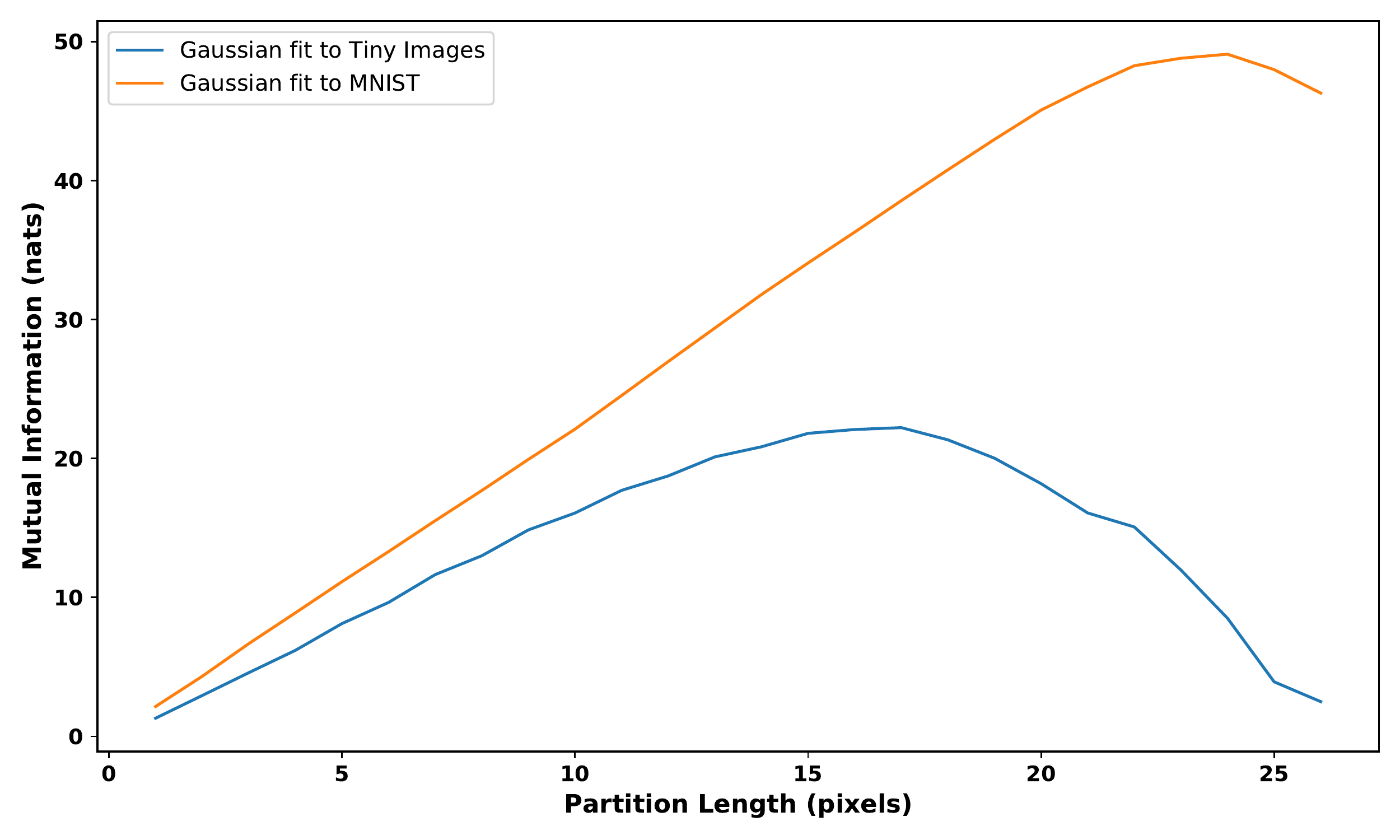}
    \caption{Analytic MI curves from the GMRFs fitted to MNIST and the Tiny Images, plotted relative to the side length $L$ of the inner partition. The Tiny Images curve shows a clear boundary law, while the MNIST curve also starts linear but gradually bends over. Since the GMRFs only model simple pairwise correlations, these MI values are very likely underestimates.}
    \label{gauss-mnist-tiny}
\end{figure}

Figure \ref{gauss-mnist-tiny} shows the Gaussian MI curves generated by fitting GMRFS to the covariance matrices of both the MNIST and Tiny Images datasets. It is important to note the scale of the y-axis: the MI values obtained from the fitted GMRFs are roughly five times larger than the predictions of the logistic regression algorithm that are shown in Figure \ref{mnist_tiny}, indicating a severe underestimation in the latter. The curve for the Tiny Images in Figure \ref{gauss-mnist-tiny} is remarkably linear, only declining at the end because of the finite size of the image. This agrees with the shape of the logistic regression curve in Figure \ref{mnist_tiny} and almost exactly resembles the boundary-law GMRF curve from Sec.~\ref{sec nearest-neighbor} (Figure \ref{area_large_small}). The MNIST GMRF curve is also approximately linear up to an inner patch length of roughly $L = 15$ pixels, at which point the curve bends over and begins to decrease due to finite size effects; these are exacerbated by the fixed black border placed around each digit\footnote{The MNIST digits themselves are only $20 \times 20$ pixels in size; since the digits are roughly centered in the $28 \times 28$ image, most of the outer pixels on the edges will be uniformly black and thus contribute nothing to the MI.}. While the MNIST curves in Figures \ref{mnist_tiny} and \ref{gauss-mnist-tiny} have somewhat similar shapes at larger patch sizes, the linearity of the Gaussian MNIST curve in Figure \ref{gauss-mnist-tiny} at small $L$ is not present in the corresponding regression curve of Figure \ref{mnist_tiny}.
Taken together, these results show that if the GMRF estimates are viewed as approximations of the simple, pairwise correlations in the images, then it is evident that the scaling behavior of those correlations obeys a clear boundary law in both datasets. Samples and covariance plots from the two fitted GMRFs are given in Figure~\ref{fig tiny_mnist_cov}.

Although the primary focus of this work is the use of logistic regression as a means of quantifying MI scaling, it is clear from Figure \ref{gauss-mnist-tiny} that GMRF techniques offer a viable alternative. We provide here a brief discussion of the relative merits of each method. Compared to a stochastically-optimized neural network, a multivariate Gaussian is very simple to fit and provides a single, deterministic MI estimate via Eq.~\eqref{eq gauss_mi}. The logistic regression algorithm, by contrast, shows significant variation across trials even when the dataset is fixed, a problem which becomes more severe at larger MI values (see, e.g., the Tiny Images plot in Figure \ref{mnist_tiny}). The simplicity of the GMRF comes at a cost, however, since Gaussians are inherently quadratic and thus incapable of modeling interactions between more than two variables. We would expect complex datasets to posses these higher-order dependencies, which favors the use of more expressive neural network models. At the same time, we can see from comparing Figure \ref{mnist_tiny} with Figure \ref{gauss-mnist-tiny} that the logistic regression method captures only a fraction of the total magnitude of the MI. Collectively, these observations suggest that the GMRF approach should be favored when the correlation patterns are simple or when only a rough lower-bound on the MI of a dataset is desired. By contrast, regression with a neural network is better suited to estimate the MI of data with more complex correlations.

\section{Discussion}\label{sec discussion}

Recent work in quantum many-body physics has shown that the success of a tensor network ansatz is closely tied to the correlation structure of the underlying system. It stands to reason that similar logic should hold in machine learning. If true, this presents us with two main challenges. First, on a theoretical level, we must gain insight into the mathematical relationships that exist between dataset correlations and network architecture. At the same time, on a more practical level, we need to be able to quantify and characterize the kinds of correlation structures present in real-world data. Our work here addresses both of these problems, using the classical MI to establish an entanglement lower-bound for probabilistic classification tasks and finding clear evidence for boundary-law scaling in the Tiny Images dataset.

On the theoretical side, we established in Sec.~\ref{sec correlation-scaling-bound} that the MI of the data features provides a lower bound on the entanglement needed for probabilistic classification of orthogonal samples by a tensor network. We showed that direct entanglement estimates, taken from the state representing the sample distribution, are artificially upper-bounded by the logarithm of the number of samples, regardless of the nature of the distribution. When the true entanglement is expected to exceed this bound, such as for data with a large number of features, a different measure of correlation such as the MI is therefore necessary. Given that the entanglement of a network with fixed bond dimension is $n\log m$ (Eq.~\eqref{eq network entangle}), an MI estimate can help determine both the connectivity of the network $(n)$ and the size of the indices $(m)$. While the lower bound should still hold approximately on samples with small overlaps, it will be useful to explore in future work whether and to what extent it is possible to generalize this bound to non-orthogonal featurizations. Additionally, there are many machine learning tasks where the ground truth cannot be expressed as a probability or modulus---e.g., regression over the real numbers $\mathbb{R}$---and which therefore fall outside of our analysis. It seems likely that the correlation structures in these tasks would still be important when choosing the right tensor network, but the mathematical relationship is not as clear as in the probabilistic cases studied here.

Assuming that the images analyzed in Sec.~\ref{sec real-data} can be mapped to tensors with minimal overlap and that therefore the bound in Sec.~\ref{sec correlation-scaling-bound} applies, then our numerical results suggest that the MI of the Tiny Images obeys a boundary law. The evidence is less definitive for MNIST, although the analytic curve obtained by fitting a GRMF shows a clear boundary law for smaller patch sizes. This would indicate that the most appropriate tensor network to use for probabilistic classification of these datasets from a correlation standpoint is PEPS, whose connectivity follows a 2-D grid. However, given that exact contraction of a large PEPS network is impossible even with small bond dimension, it would be useful to look at alternative structures that still possess a 2-D geometry. Some possibilities include a TTN with four child nodes, or networks with a Cayley tree structure~\cite{Li_von_Delft_Xiang_2012} possessing four nearest neighbors.

From a numerical perspective, our present work on MI estimation appears to be one of the few in the literature that seeks to quantify the spatial structure of the MI, or even just approximate the magnitude of the MI itself. Instead, most of the existing research focuses on MI as a minimization or maximization target, as seen in various independent component analysis algorithms~\cite{Kong_Vanderburg_Gunshin_Rogers_Huang_2008} or in the training of generative models~\cite{Chen_Duan_Houthooft_Schulman_Sutskever_Abbeel_2016}. To our knowledge, the only other work that explores MI scaling is that of Cheng et al.~\cite{Cheng_Chen_Wang_2018}, which characterized the MI of MNIST in the context of training sparse Boltzmann machines. The authors utilized side-to-side and checkerboard partitioning schemes, focusing their analysis on the degree to which the estimated MI value (using Kraskov's nearest-neighbor method) differed from the maximum MI value that could exist between the partitions. While their results showed that the estimate was significantly smaller than the maximum, it is unclear how much of this was actually an intrinsic property of the data or just a numerical limitation of the nearest-neighbor method used for estimation.

Indeed, recent work by McAllester and Stratos~\cite{McAllester_Stratos_2020} has shown that lower-bound MI estimates based on sampling, such as our logistic regression algorithm using the DV representation, can never produce an estimate greater than $\mathcal{O}(\log N)$, where $N$ is the number of samples. If we make the reasonable assumption that the Gaussian curves from Figure \ref{gauss-mnist-tiny} underestimate the true MI, then we would need on the order of $10^{21}$ images to get a good estimate of the Tiny Images MI. This is of course impossible. For MNIST, the number of samples needed is on the order of $10^8$, which is within the realm of possibility but would require a massive data collection and training scheme. On a practical level, this means that the DV representation cannot be used for MI estimation on datasets that have strong correlations, although it is unclear whether the $\log(N)$ bound tells us anything about direct approximations of the KL divergence in the spirit of Eq.~\eqref{eq direct-est} (which was used to produce Figure \ref{mnist_tiny}). McAllester and Stratos recommend instead to minimize the cross-entropy as an upper bound on the entropy, then use Eq.~\eqref{eq13} to get an estimate of the MI that is not a lower bound. This could be a useful direction for future work.

Tensor network machine learning is still in its infancy, and there is much work to be done in understanding the strengths and weaknesses of different network designs. It is likely that dataset correlations present in a given task will dictate the tensor structure that is best suited for the job, but determining which correlations are most important, and knowing how to assess that importance, is challenging. We have shown here that the scaling of the MI within a dataset can be systematically characterized in a manner that parallels the entanglement scaling analysis performed on quantum states, which may provide insight into these questions.

\section*{Acknowledgements}

I.C. was supported by the US Department of Energy, Office of Science, Office of Advanced Scientific Computing Research, Quantum Algorithm Teams Program, under contract number DE-AC02-05CH11231. W. H. was supported by a grant from Siemens Corporation. H. L. was supported by the National Aeronautics and Space Administration under Grant/Contract/Agreement No.80NSSC19K1123 issued through the Aeronautics Research Mission Directorate. Computational resources were provided by the Molecular Graphics and Computation Facility at the University of California, Berkeley (NIH grant S10OD023532).

\printbibliography

\section{Appendix}\label{sec appendix}

\subsection{MNIST and Tiny Images Datasets}\label{sec appendix_data}

\begin{figure}
    \centering
    \includegraphics[width=\textwidth]{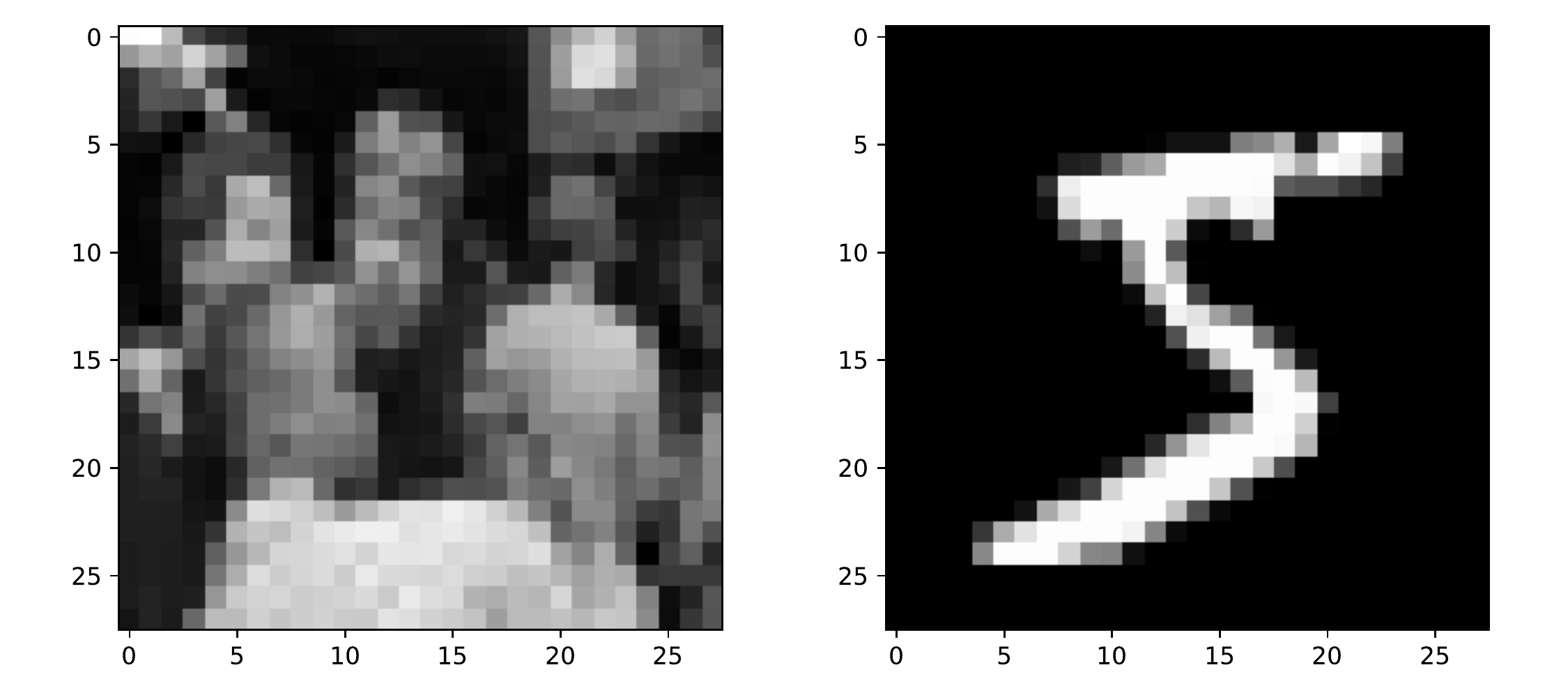}
    \caption{Left: Low-resolution image taken from the Tiny Images dataset after having been cropped and converted to grayscale. Right: The handwritten digit ``5" taken from MNIST. The axes give the height and width of each image.}
    \label{tiny-mnist}
\end{figure}

MNIST and the Tiny Images datasets (Figure \ref{tiny-mnist}) are common benchmarks used in computer vision research. MNIST consists of 70,000 samples of handwritten digits collected from high school students and Census Bureau employees by the National Institute of Standards and Technology (NIST) and further processed by LeCun, Cortes, and Burges. The original NIST images were bilevel, with each pixel represented by a single bit as either black or white. Grayscale shades were then introduced incidentally when the digits were reshaped to fit in a $28 \times 28$ array.

The Tiny Images dataset is a set of approximately 80 million images collected by Torralba, Fergus, and Freeman. The dataset was gathered from the internet by searching for $75,062$ nouns using a variety of search engines. The images were downsampled to $32 \times 32$ pixels, with each pixel represented as a vector in RGB color space. For a better comparison to MNIST, we converted the colored images to grayscale and cropped them to down to a size of $28 \times 28$. 

\subsection{GMRF Covariances and Sample Images}\label{sec appendix_gmrf}

\begin{figure}
    \centering
    \includegraphics[width=\textwidth]{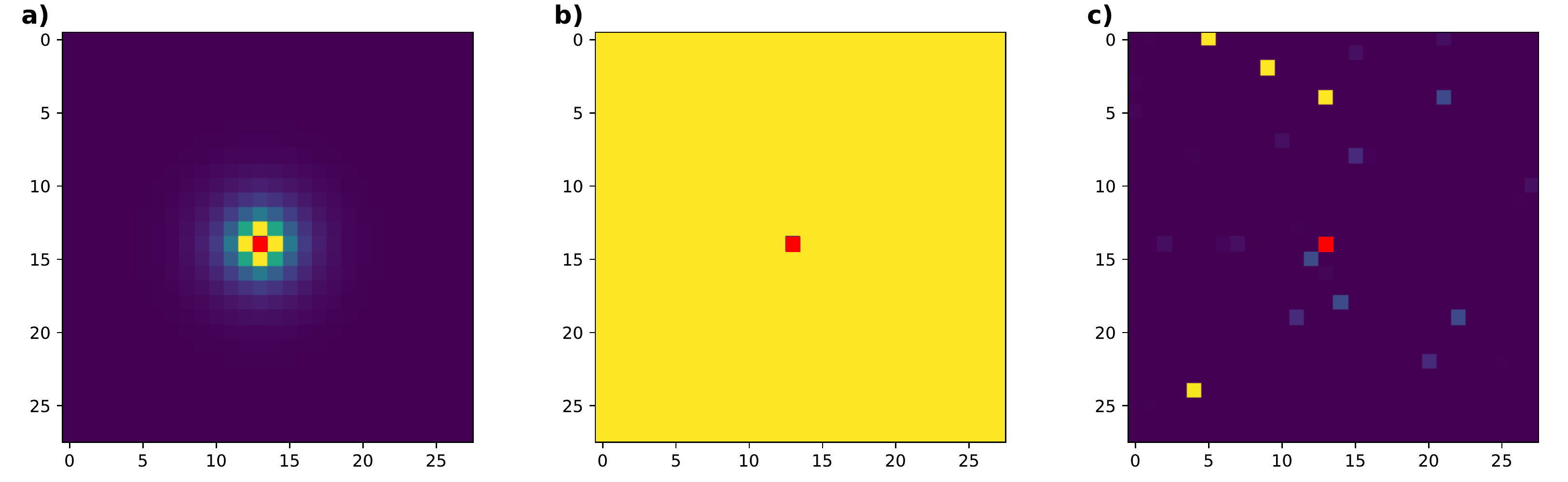}
    \caption{Covariance plots for the \textbf{a)} nearest-neighbor, \textbf{b)} uniform, and \textbf{c)} randomized GMRF distributions used in Sec.~\ref{sec GMRF}. The covariance values are taken with respect to the center variable highlighted in red, with brighter colors indicating stronger correlations and darker pixels indicating weaker correlations.}
    \label{gmrf_cov}
\end{figure}

\begin{figure}
    \centering
    \includegraphics[width=0.9\textwidth]{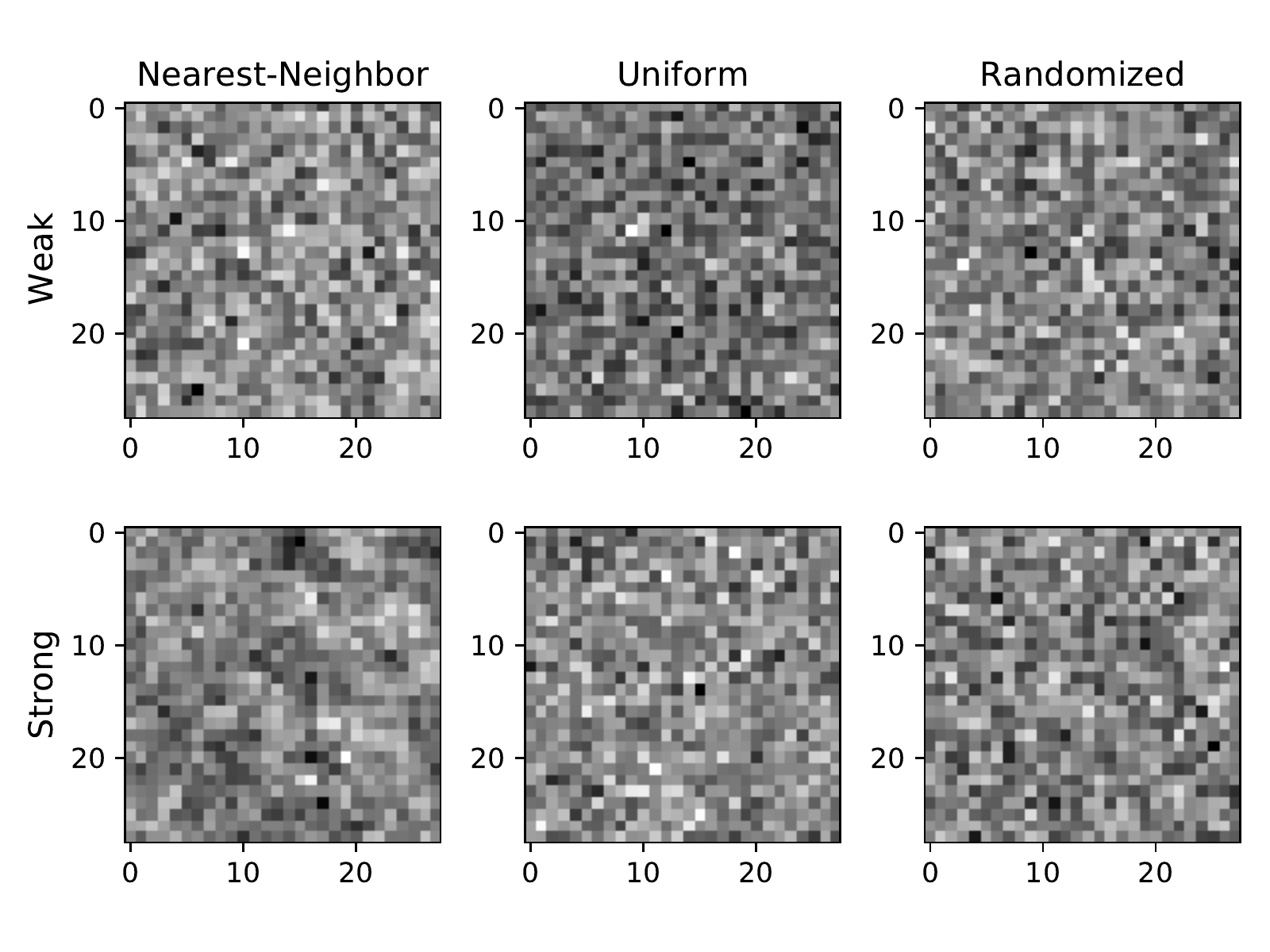}
    \caption{Sample ``images" taken from the GRMF distributions of Sec.~\ref{sec GMRF} at both strong and weak correlation strengths.}
    \label{fig grmf samples}
\end{figure}

Figure~\ref{gmrf_cov} shows covariance plots of the three GMRFs tested in Sec.~\ref{sec GMRF} with respect to a single variable highlighted in red. The magnitudes are expressed as colors to emphasize the importance of the correlation pattern rather than the specific covariance values. The variables that have the strongest covariance with the center variable are bright yellow, and correspond to the variables which have a non-zero conditional correlation with the center variable. In Figure~\ref{gmrf_cov}a the four nearest-neighbor variables are clearly visible, while in Figure~\ref{gmrf_cov}c those four variables are randomly distributed throughout the image. In Figure~\ref{gmrf_cov}b the covariance matrix is uniformly yellow, as every variable is conditionally correlated with every other variable.
Samples from these GRMFs are shown in Figure~\ref{fig grmf samples}, where the subtly of the correlation effects is evident. 

\begin{figure}
    \centering
    \includegraphics[width=0.7\textwidth]{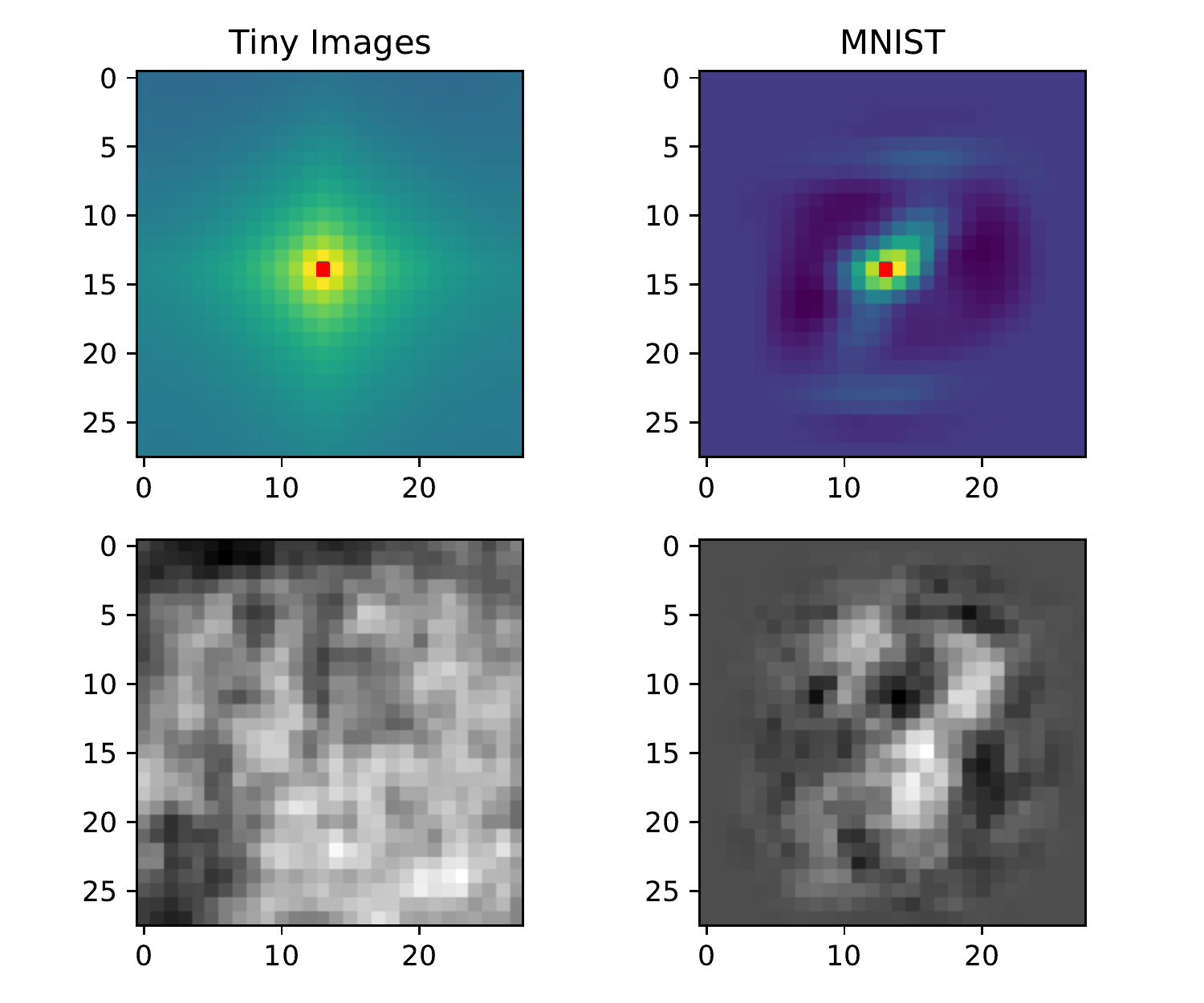}
    \caption{The covariances (top row) and sample images (bottom row) from GRMFs fit to the Tiny Images and MNIST datatsets. The covariance values are calculated with respect to the central pixel highlighted in red, with brighter colors indicating larger values. The Tiny Images covariance plot shows a strong nearest-neighbor pattern, while the MNIST plot has a more complicated and long-range structure. The sample images show some structure, but are not identifiable as a digit or object.}
    \label{fig tiny_mnist_cov}
\end{figure}

The covariance plots and sample images shown in Figure~\ref{fig tiny_mnist_cov} are taken from the GRMFs fit to the Tiny Images and MNIST. The samples posses considerably more structure than those in Figure~\ref{fig grmf samples}, which is consistent with the large MI values found in Figure~\ref{gauss-mnist-tiny}. That said, the GRMFs are clearly not able to capture the full structure of the underlying dataset distributions, since the Tiny Images GMRF does not resemble any identifiable object and the MNIST GMRF sample does not resemble any digit. The covariance plots of Figure~\ref{fig tiny_mnist_cov} both show strong nearest-neighbor correlations, which is consistent with the boundary-law scaling observed in Figure~\ref{gauss-mnist-tiny}.  

\end{document}